\documentclass[a4paper,12pt]{article}
\usepackage[pdftex]{graphicx}
\usepackage{amsmath}
\usepackage{plain} 
\usepackage{url}
\usepackage[utf8]{inputenc}
\usepackage{longtable}
\title{Intergroup violence in bursts}
\author{Jeroen Bruggeman\thanks{Corresponding author: Jeroen Bruggeman, \texttt{j.p.bruggeman@uva.nl}}, Don Weenink, Bram Mak}
\date{ }
\DeclareMathOperator{\arcosh}{arcosh}
\begin{document}  
\maketitle


\begin{abstract}

During intergroup confrontations, agitating stimuli such as opponents' threats and provocations can trigger collective violence, even without the usual mechanisms of ingroup cooperation, such as norms with sanctions. We examine video recordings of street fights between groups of young men. Their violence sometimes breaks out in a burst, wherein a majority of participants starts fighting almost simultaneously. At other times, only few group members participate and it takes them more time to do so. This difference in commencing collective violence can be understood by perceiving it as a collective action dilemma. We adapt an Ising model to show that the proportion of group members who cannot or do not want to contribute to the public good---victory over opponents---predicts whether violence takes the form of a burst or not.
 
\end{abstract}

\section{Introduction}
In studies of intergroup conflict, the large numbers of casualties of wars between and within countries draw most attention \cite{clauset20,richardson48}. Yet, the most frequent occurrences of intergroup violence involve small groups, including subgroups of larger groups. We focus on small groups, and want to explain how violence committed by unarmed, non-professionals starts off.
When observing pertinent videos, one often notices bursty starts of violence, where the majority of a group engages almost simultaneously in violence against their opponents. Before punches are thrown, individuals in the focal group and their opponents insult and threaten, and thereby agitate, one another. For this to result in collective violence by focal group members, a critical level of agitation has to be reached. A necessary condition is that focal group members are in close proximity and maintain face-to-face contact \cite{collins08,cialdini04,mcdoom13}.  

In one of the most influential studies of violence, Randall Collins \cite{collins08} argues that antagonists need to circumvent an emotional barrier of tension and fear in order to be able to use violence. Surmounting the barrier is easier when the other party appears vulnerable or weaker, or when the focal group is supported by an audience \cite{collins08}. Collins’ notion of ``forward panics" corresponds to the bursts we observed. In forward panics, assailants seize the opportunity to forcefully attack opponents in moments of vulnerability or weakness, for example when being isolated or falling down to the ground. This emotional barrier is meshed with another barrier known as the dilemma of collective action \cite{olson65,ostrom09}, which has gained less attention in scholarly work on violence. We will present a formal model that explicates the conditions for groups to burst into collective violent action. At the core of the model is opponents' agitation. 

When agitation passes a critical level (for example, when one individual is pushed by an opponent), violence can break out in a burst. Alternatively, the confrontation can escalate more slowly and with fewer participants. Understanding how collective violence starts is complicated by the presence of other people, who may form an audience or try to de-escalate \cite{levine11,phillips05,weenink22}. De-escalation happens often in street fights \cite{philpot20}. Given the conditions of proximity, agitation, and de-escalation, our goal is to explain bursts versus non-bursts. To this end we use the centennial Ising model from physics \cite{macy24,stein13}. It resulted in a Nobel Prize for Parisi (in 2021; \cite{parisi06}) and has been used in various studies of social influence \cite{castellano09,stauffer07,jones85,galam91}. We adapt the model in a novel way, use it as an agent based model, and apply it to videos of street fights between groups of (mostly) young men. This work stands in a tradition of statistical physics models of collective behavior, including violence and its preludes, such as coalition forming \cite{vinogradova13} and polarization \cite{baumann20}. Our Ising model has in common with models of social contagion \cite{watts02b,ruan15} that individuals' behavior (or opinion) depends on their network neighbors. Furthermore, bursts of violence resemble the outbreak of synchronization in the Kuramoto model \cite{arenas06}. In conflicts at different scales, both event sizes (number of victims) and interevent times are approximately power law distributed \cite{bohorquez09,johnson11,johnson13,okamoto23,picoli14}. Below, we will first provide an informal overview related to social science theory before we proceed with the formal Ising model, followed by our empirical study.    

\section{Theory}
Collective violence involves a dilemma of collective action \cite{ostrom09,olson65}, so before predicting bursts, we must first address how the Ising model explains overcoming such dilemmas. For an unarmed focal group, the ingroup ties that matter are face-to-face contacts in close proximity \cite{collins08}. We already know from public goods experiments that if participants regard a public good to be valuable, more than half of the participants are conditional cooperators willing to contribute if many others do \cite{chaudhuri11}. In other words, conditional cooperators conform to their (weighted) average neighbor in the network \cite{blau64,toelch15}. 
In our case, the public good is victory over opponents, which means that opponents flee or are wrestled to the ground. If victory is not feasible, a second tier public good is the prevention of ingroup humiliation. 
When confronted with opponents, most individuals experience uncertainty, which increases the likelihood of their conforming to the group \cite{wu14,morgan12}. Conformity makes sense from an evolutionary perspective when payoffs are hard to predict \cite{vandenberg18}. Therefore, individuals' motivation not only depends on their appraisal of the public good, but also their desire to align with their group members.

In the simplest case, there are two individuals, each with the behavioral options to defect (not fight) or cooperate (fight). The possibilities are: (1) both individuals defect, which avoids exploitation but does not yield any public good; (2) one individual cooperates at a cost, yielding half of the public good, but is exploited by the freeriding other; or (3) both cooperate at a cost, which maximizes the public good. At the beginning, both defect. Without additional motivation, the probability that one of the two will start contributing to the public good is very small, due to high risk and small yield. By contrast, if one of the two cooperates, the other conditional cooperator is likely to join in. Yet, it is very unlikely that a group of non-fighting individuals ($n \geq 2$) will act collectively, and even less likely in larger groups where the effect of shirking is less noticeable \cite{olson65}. 

This equilibrium at defection can be perturbed by the turmoil of confrontation, which we define as opponents' threats. Threats agitate members of the focal group, who produce the hormones norepinephrine and cortisol that linger on for a while. Alternatively, turmoil may signal opportunity, for example opponents who stumble or become isolated from their group. Due to turmoil, one or few individuals in the focal group may accidentally cooperate even when most others do not (yet). This accidental or spontaneous cooperation is caused by norepinephrine and the emotional responses to turmoil and is called ``trembling hands'' in game theory \cite{dion88}. Among conditional cooperators, a few trembling hands can influence proximate others and may initiate a cascade of cooperation, or a burst if the cascade proceeds rapidly.  

In the model, opponents' turmoil (threat or opportunity) has an effect on the focal group at the aggregate level and affects everyone equally. In larger groups than ours, individuals standing closer to opponents are exposed to more turmoil than others standing further away. Locally varying turmoil in the Ising model as well as network variations are examined elsewhere \cite{bruggeman20a}. Here, we make the model not more complicated than necessary for our empirical study. The model demonstrates that if turmoil increases, there is no gradual increase of accidental cooperators. Instead, when turmoil reaches a critical level, fighting breaks out in a burst. Bursts clearly differ from a gradual build up of collective action and the distinction does not follow from other models in the literature. One might expect, by contrast, that the tipping point is smooth rather than sudden (it is a second order phase transition \cite{nishimori01}), and in large random networks (e.g., $n = 1000$) it is. However, all large social networks are decomposed into smaller clusters, where the transition is sudden rather than smooth, also in time (over Monte Carlo steps); in small clusters, the tipping point occurs at lower turmoil and lower levels are reached earlier on \cite{bruggeman20a}. 

The distinction between bursty and slowly increasing collective action turns out to depend on the critical proportion of non-fighters (versus conditional cooperators) in the focal group. Perhaps they would be conditional cooperators under different circumstances, but they were not in our study. Hence, we calculate this proportion and use it as a prediction that we test empirically. In violent situations, many people do not fight, for all kinds of reasons. They may be too scared to fight \cite{collins08}, have empathy with their opponents, disagree with violence (i.e.,~appraise the public good differently), feel no solidarity with their group \cite{collins04}, try to de-escalate, be stopped in their tracks by de-escalators (also called guardians in criminological studies), or they may have fought but were wounded, were forced to the ground, or became exhausted at some point and became passive. In the population at large, preferences for violence will have a distribution \cite{fischbacher06} with a negative mean value that expresses a general dislike of violence. Only few individuals in the right tail of the distribution will self-select into violent groups and value the public good positively.\footnote{Milliff \cite{milliff23} suggested that these people perceive their situation as uncertain but controllable.}   
If the proportion of non-fighters in these groups stays below the critical level, turmoil will be followed by a burst of violence. Above the critical level of non-fighters, however, there is neither a critical level of turmoil nor a burst. 
Note that the non-fighters incorporate not only de-escalators in the focal group, but also focal group members who have been prevented from fighting by de-escalators in the focal group or by others.  

Most models of collective action in the literature revolve around individual rewards and punishments, called selective incentives \cite{olson65}, on top of a share of the public good. These incentives require norms about (in)appropriate behavior in a given situation \cite{fehr03}, as well as monitoring of group members \cite{rustagi10} and transmission of information (i.e., gossip) through the group's network, which leads to reputations \cite{nowak05} that feedback through selective incentives, with or without leaders. This package of mechanisms is crucial for ongoing cooperation in the longer run but needs time to develop, which may not be available when threats are imminent. 
The more time people have, the better they can prepare themselves, which is especially important for high-risk situations. Examples of well-prepared groups are police, soldiers, firefighters, and combat medics who receive professional training that enables them to cooperate effectively and respond to situational stimuli in predetermined manners rather than spontaneously.  By contrast, ordinary citizens and amateur fighters lack training and team coordination. For non-professional groups, such as ours, the uncertainties of outcomes, benefits, and costs are higher. 

Several earlier models of cooperation do not feature the usual package of cooperation mechanisms; namely, models of thresholds \cite{granovetter78}, cascades \cite{watts02b}, and critical mass \cite{marwell93}. Therefore, these models are potentially useful for our study. They draw on the assumption that some people take the initiative or leaders set cooperation in motion. Yet in many cases, violence occurs in the absence of leaders \cite{ives20}. The Ising model points out how cooperation can start spontaneously without leaders, galvanized by accidental cooperators rather than exceptionally zealous ones. If there are initiative takers or leaders in favor of violence \cite{glowacki22}, however, they can be accommodated in the model. 

These earlier models also draw on strong rationality assumptions, such as perfect information about the numbers of cooperators at every moment. Yet, in violent and other uncertain situations, the degree of rationality is bounded due to incomplete or absent information about opponents and group members beyond face-to-face contact. For example, these models ignore opponents' moments of vulnerability, which are important in actual conflicts and augment agitation in the Ising model. Moreover, individuals often respond to provocations in ways that harm their interests, by overestimating their own abilities and underestimating the tenacity of their opponents. The Ising model is more parsimonious than the aforementioned models because it does not rely on assumptions of strong rationality. For the model it is enough if individuals find the public good valuable, without the researcher assumes how (in)accurate their expectations are. The model combines the precision of game theory, including a mapping on payoffs---usually unknown to the participants---with the behavioral spontaneity of chaotic and uncertain situations \cite{steinert17}. Beyond threshold models, this uncertainty has been assessed qualitatively \cite{snow14} instead of modeling it precisely \cite{macy91}. Uncertainty has been added to threshold models, too, which makes it possible to loosen the rationality assumptions. Then, a small probability, $\epsilon$, that an individual starts cooperating by random chance suggests that the probability that someone in a group cooperates increases exponentially with group size, $1 - (1-\epsilon)^n$ \cite{macy15}. However, this argument omits the interdependence of group members. Consequently, it contradicts our empirical finding that cooperation starts earlier in smaller groups, as predicted by the Ising model. Finally, whereas game theoretic approaches to conflict readily become complicated, even if there are only two individuals \cite{shadmehr11}, the simplicity of the Ising model makes it possible to investigate large networks computationally. 


\section{Model}
\begin{figure}[!ht]
\begin{center}
\includegraphics[width=.55\textwidth]{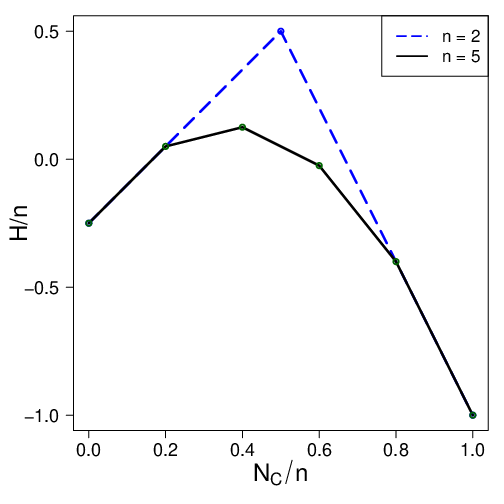} 
\end{center}
\caption{The dilemma of collective action presented as a hill between full defection (left) and full cooperation (right), with the proportion of cooperators $(N_C/n)$ on the horizontal axis. Data points are based on Equation~\ref{eq:ising}, for $C$ = 1 and $D$ = 1/2.  The vertical axis $(H/n)$ could be intuited as a negative likelihood; moving uphill from a state of defection $(N_C/n = 0)$ is very unlikely. One line (dashed) is drawn for a dyad and one for a clique of five individuals. The larger the group is, the more rounded the hill becomes.} 
\label{fig:mountain}
\end{figure}

Members of a (possibly fledgling) focal group, indexed $i$ or $j$, can defect, $D$, or contribute, $C$, to a public good, with $0 < D < C$.  Behavioral variable $S_i$ of conditional cooperators can take the value $ S_i=C $ or $ S_i=-D $, whereas non-fighters stay put at $S_j=-D$.  
The behavioral options correspond to magnetic spins in the original Ising model. Therein, turmoil, $T$, is temperature.
Before a collective action, at low $T$, everyone in a focal group defects. Network ties among focal group members, $A_{ij} > 0$, mean that $i$ is in close proximity to and senses the behavior of group member $j$, or else $A_{ij} = 0$.  Because people tend to respond to proportions of their social environment rather than absolute numbers \cite{watts02b,friedkin11}, ties are row-normalized, with $w_{ij} = A_{ij}/\sum_{j=1}^n A_{ij}$ such that $\sum_{j=1}^n w_{ij} = 1$. Based on our video data, we assume that in our small groups ($n < 10$), sensing is reciprocal at least to some degree, but not necessarily symmetrical, and that everyone senses all other group members, thus the network is fully connected.\footnote{Temporary exceptions to full connectedness in our data were groups where some people's view was blocked by objects, opponents, or bystanders, as well as one larger group ($n=14$). These cases we investigated through simulations.} 
The Ising model is the following Hamiltonian equation \cite{barrat08,walter15} 
\begin{equation}
H = - \sum_{i \neq j}^{n} w_{ij} S_i S_j.
\label{eq:ising}
\end{equation} 
Fig.~\ref{fig:mountain} plots $H$ with the normalized number of cooperators, $N_C/n$, from left to right.

We do not assume that individuals know their payoffs in advance, but they will heuristically---and perhaps wrongly---distinguish between valuable ($C > D$) and non-valuable ($C < D$) public goods or, to the same effect, between potentially efficacious and non-efficacious actions \cite{saab16}. Note that payoffs are not used in the model's calculations but are defined to provide a meaningful interpretation. When $i$ chooses between $C$ and $D$ amid $N_C$ cooperators, payoffs for cooperation and defection are, respectively 
\begin{align}
P_C &=\theta(N_C + 1)/n - 1 \\
P_D &=\theta N_C/n + Q 
\end{align}
with a synergy or enhancement factor $\theta \geq 1$. These definitions are the same as in evolutionary game theory \cite{perc17} except for $Q$. This additional factor $Q$ assures that if $D$ approximates $C$, which means that the outcomes of defection and cooperation become equally valuable, $P_D$ approximates $P_C$.\footnote{$Q= (\theta/n - 1)(1-R); R = (C-D)/(C+D); \theta = \theta_0 + R$, with a base rate $\theta_0 \geq 1$.} 
If one fine-tunes $C$ and $D$ in $Q$ to increase or decrease the difference between $P_D$ and $P_C$, this will alter the difference in height between the two minima in Fig.~\ref{fig:mountain}.

For our empirical study, we have to choose values for $C$ and $D$ to predict the critical proportion of non-fighters, $p_c$. The most obvious choice is $C = 1$, as in game theory and the original Ising model. For $D$ we want to avoid two trivial values: if $D = 0$, there is no dilemma (but a downward slope to the right in Fig.~\ref{fig:mountain}), and if $D = 1$, there is no point in cooperating, which is equally valuable as defecting. Choosing $D$ to be maximally distant from the two trivial values seems to be a reasonable first approximation. Hence, we set $S = \{1,-1/2\}$ for all conditional cooperators, also in the examples. 
For non-fighters $j$ we set $S_j = - 1/2$, irrespective of their reasons. Note that all earlier Ising models had either the values $\{1,-1\}$ \cite{castellano09,stauffer07,galam91,weidlich71} or $ \{1,0\}$ \cite{daniels17}. Of all these models, only one represents a public goods game, in that case for two individuals \cite{adami18,sarkar19} whereas our model is applicable to groups of any size.   

Beyond our empirical study, the payoffs in the asymmetric Ising model can be generalized by relating $C$ and $D$ to the symmetric model through a mapping $\{C, -D\} \rightarrow \{ S_0 + \Delta, S_0 - \Delta \}$, with a bias $S_0 = (C-D)/2$ with respect to 0, and the two behavioral options symmetrical at each side of $S_0$ at an offset $\Delta = \pm (C+D)/2$. It can be shown that the asymmetry in $S$ is equivalent to the symmetric model with an external field $2 S_0$ \cite{brugspin}.   
In the payoffs, the bias and offset are expressed through $R = S_0/\Delta$. If $\Delta$ is set to a fixed value (here 0.75), decreasing $S_0$ (here 0.25) makes the public good and cooperation for it less valuable. This decline is equivalent to an increasing proportion of cooperating network neighbors that is necessary to win over an actor to cooperate, also in other binary decision and contagion models \cite{watts02b,granovetter78}. Increasing $S_0$ makes cooperation more valuable, and it corresponds to a decreasing proportion of cooperating network neighbors. If there are initiative takers, indexed $j$, they will have a higher $S_{0,j}$ than the other group members, which lowers the critical turmoil level, $T_c$, in small networks.

\begin{figure}[!ht]
\begin{center}
\includegraphics[width=.4\textwidth]{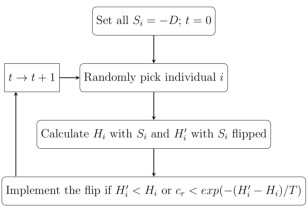} 
\end{center}
\caption{The Metropolis algorithm. Index $i$ runs over the conditional cooperators, not the non-fighters; $t$ is a counter of Monte Carlo steps; and $c_r$ is a random number in the range $0 \leq c_r \leq 1$.} 
\label{fig:algorithm}
\end{figure}

Solving the Ising model boils down to minimizing $H$, which can be done analytically by a mean field approach (in the Supplementary Material). We prove that if $C = 1$ and $D = 1/2$, $p_c = 1/3$. For inhomogeneous (i.e., clustered) networks, the mean field approach is inaccurate, but $H$ can also be minimized computationally, which is simpler and can deal with inhomogeneous networks.
The computational approach over a certain range of $T$ is as follows (Fig.~\ref{fig:algorithm}). For a given level of turmoil, a network node $i$ is randomly picked and $H$ is calculated. For comparison, $i$'s current behavior is flipped from $D$ to $C$ (or the other way around if $i$ cooperates) and $H'$ with the flip is calculated. The flip is accepted and implemented if $H' < H$ or with a certain probability that increases with $T$. In other words, increasing $T$ increases the amount of randomness (due to turmoil) in $i$'s decision. A behavioral change of $i$ affects others in $i$'s neighborhood when it is their turn to decide. Consecutive decisions are Monte Carlo steps in the Metropolis algorithm \cite{barrat08} that loops through great many Monte Carlo steps (counted by $t$ in Fig.~\ref{fig:algorithm}) in order to allow individuals' interdependent behavior to settle down. This procedure is repeated at increments of $T$, in our study with step size 0.01. 

\begin{figure}[!ht]
\begin{center}
\includegraphics[width=.55\textwidth]{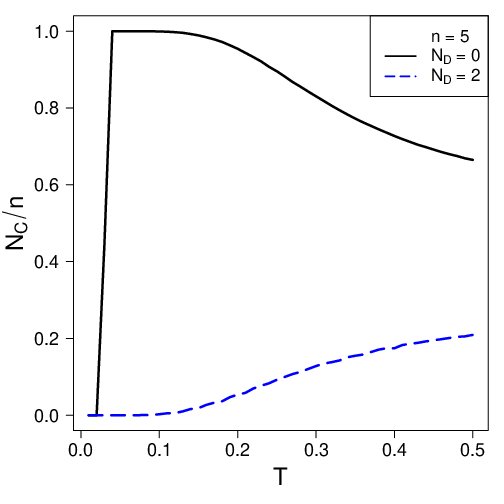} 
\end{center}
\caption{Simulation of Equation~\ref{eq:ising} over a range of turmoil ($T$) on the horizontal axis. The proportion of cooperators ($N_C/n$) in a fully connected network in the size range of our empirical study ($n = 5$) is on the vertical axis; $C = 1$ and $D = 1/2$.  The black line (top) depicts a burst in the group without non-fighters ($p = 0$). The dashed blue line (bottom) depicts a gradual increase of cooperation in the group with two non-fighters ($ p > p_c$).} 
\label{fig:burst}
\end{figure}

To illustrate the emergence of cooperation with increasing turmoil in a group of $n=5$, we apply the Metropolis algorithm; see Fig.~\ref{fig:burst}. First, we investigate the case without non-fighters. At low turmoil, collective action does not start (continuous line), but at a critical level $T_c$, (almost) everybody bursts into cooperation, with a maximum (where $N_C/n \approx 1$) close to $T_c$. Then, a small number of accidental cooperators wins over most others to join the collective action. The effect of turmoil is nonmonotonic and the level of cooperation decreases if $T$ keeps increasing beyond $T_c$, which means that extraordinarily strong turmoil becomes more confusing than motivating to fight. 

If there are non-fighters, $T_c$ increases, which is more pronounced in larger networks, and maximum cooperation is lower than in the previous case because the number of conditional cooperators is lower.\footnote{In Fig.~\ref{fig:mountain}, a proportion of non-fighters, $p$, means that no matter how many conditional cooperators contribute, $N_C/n \leq 1-p$.} 
If the proportion of non-fighters surpasses a critical level, $p_c$, and $T$ increases above $T_c$, there is no burst (dashed line in Fig.~\ref{fig:burst}) and cooperation increases gradually to a low level. In the mean field analysis, there is no (burst of) cooperation if $p_c > S_0/\Delta$, independent of network size and density. In simulations, $p_c$ is proportionally less precise in smaller networks (Table~S3), but because our networks are complete, we stick to the mean field prediction. Based on the mean field analysis ($p_c = 1/3$) and a nearly identical result from the simulations ($p_c = 0.34$),\footnote{If non-fighters are clustered together in a sparse network, they are less in the way of collective action of the remainder network (thus $p_c$ is higher) than if they are evenly spread out across the network.} 
our main prediction is that $p_c = 1/3$. 

The critical threshold of turmoil increases with network size at a decreasing rate \cite{bruggeman20a}, but it also increases with the proportion of non-fighters. At $T_c$, simulations point out that cooperation starts in small clusters of conditional cooperators, which we will also test empirically. The onset in small groups is puzzling because larger groups have a better chance to win at lower individual costs. Yet if in a small (sub)group, someone starts fighting, he (and rarely she) accounts for a relatively sizeable proportion of his neighbors' social contacts, and more readily wins them over to fight than in a large group where he would comprise a small minority. 
Fighting ends when exhaustion sets in, one party dominates (i.e., opponents flee or are wrestled to the ground), or others intervene. 

\section{Data and Methods}
In studying violence, lab experiments lack the turmoil and emotional intensity of violent confrontations due to their obligation to meet ethical standards. Empirical field studies, in contrast, cannot be based on a random sample of participants or groups, yet they are invaluable to obtain a realistic view of violence \cite{bar04}. We obtained 42 videos from websites such as YouTube, LiveLeak, and WorldStarHipHop using search terms with the English keywords ``brawl,'' ``street fight,'' and ``assault.'' This sample is not random with respect to violence, but it is random with respect to temporal unfolding and (sub)group size. Of these clips, 36 are from English-speaking countries (mainly the US and the UK, with one from Canada and one from India); five of the remaining clips are from the Netherlands, and one is from Colombia. We did not observe differences in relevant behavior related to the location of the recording.
To keep distracting factors away from our analysis, we excluded clips with professional fighters, long range weapons, protective clothing, a referee, ambush attacks, or youths in a school yard. 
Most of our videos are recorded on phones by bystanders and are left-truncated; in all likelihood, there would have already been some turmoil that motivated bystanders to start filming. 
The shortest lasted 30 sec.~and the longest was nearly 5 minutes (mean 101 sec.; s.d.~59 sec.). Out of a potential 2 x 42 groups, where the opponents in one analysis become the focal group in the next, 25 groups attacked a single individual rather than a group. Because a lone individual is unable to act collectively, this leaves 59 groups to examine; see the Supplementary Material for all case studies and elaborate examples of interpreting and coding. One group had 14 members, but all other groups were small, $2 \leq n < 10$ (mean 3.6). They were simulated as cliques wherein everyone could sense one another unless there were obstacles or de-escalators obstructing contact.  

The videos were coded using Noldus Observer XT 14 software. Clips were played at half speed many times over, and one of us discussed the coding of each with one or two assistants. The assistants were unaware of the theoretical expectations. 
Each of the 406 individuals was coded for group belonging, and their behavior was interpreted and coded on the timeline.  

We coded \textit{violence} when force was used against another’s body (punching, slapping, kicking, hitting, stomping) and/or when another person’s body was forcefully moved. People may defect (not use violence) for all kinds of reasons and due to various causes. To contrast non-fighters from conditional cooperators, we coded them as \textit{non-fighters} if others were fighting while they were not. Among them we include de-escalators, as well as group members who were prevented to fight by de-escalators (who in turn might belong to either group or none).  

We subsumed the following behaviors of members of the opponent group under \textit{turmoil} for the focal group: aggressing (including fighting gestures); pulling off clothing (jackets or vests); pulling up pants (signaling readiness to fight); pointing toward opponents; provocative gesturing with fingers or hands (as an invitation to engage); bending forward toward an opponent; approaching the focal group; encroaching (invading opponents’ personal space through using or damaging objects belonging to them); teasing, such as lightly hitting or ridiculing; and violence.  
Because stumbling and falling signal vulnerability of opponents, which tends to agitate focal group members \cite{collins08,nassauer19,weenink14}, we included these mishaps in our measure. It is likely that over relatively brief time intervals, the effects of turmoil accumulate. Therefore we calculated the level of turmoil from the beginning of the video until a focal group’s (first) maximum participation in violence. We did this by multiplying the duration of each instance of turmoil by the number of individuals involved; we added up all these weighted instances. 

Given the distraction caused by turmoil, it is not feasible for all group members to react within 1 second to a group member who initiates violence (as they might in a well-organized sports team), whereas 3 seconds is too long for the notion of burst to apply to our small groups. Hence we defined a \textit{burst} as an outbreak of violence by at least half of the group (Fig.~\ref{fig:burst}), or both individuals in a dyad, if they started fighting less than 2 seconds after the first, with a 5\% margin.\footnote{The requirement that $N_C \geq 0.5 n$ is based on simulations of small networks just above $p_c$; at this proportion in large networks (e.g., $n = 1000$), $N_C < 0.5 n$.} 

Our data do not enable us to assess causality. Instead, we assess if the patterns in the data support or refute the theoretical predictions.

\subsection{Ethics} 
The use of videos for research purposes poses distinct ethical challenges, largely due to the non-anonymous content of the videos. However, ethical guidelines for digital spaces tend to be less restrictive \cite{milne12}, with the consent of the participants being less stringent for data acquired from the public domain, including the internet. While our video corpus is open for use and inspection by other researchers upon request, we require that they take the same measures to ensure the anonymity of the persons portrayed as we did.  

\section{Results} 
Of the 59 groups considered, there were 23 groups where violence started in a burst, 15 groups where violence was collective without a burst, and 21 cases of violence by a single group member. The distributions of violent event and interevent durations is presented elsewhere \cite{bruggeman26b}. Turmoil preceded all collective violence with one exception, where two individuals suddenly assaulted a passive victim. The critical level of turmoil ($T_c$) for bursts is case-specific and depends on group size, both in absolute number and relative to the size of the opponent group, and on the proportion of non-fighters. Additionally, the use of weapons has an intimidating effect.   

We confirm the finding of an earlier study \cite{collins08} that fighting tends to start in small groups or in small subgroups of larger groups. As predicted \cite{bruggeman20a}, we found that small (sub)groups start fighting at lower levels of turmoil than larger groups, and lower levels are of course reached earlier. The local emergence (versus central coordination) of collective action in small groups has also been observed in protests \cite{steinert17} and is similar to bottom-up synchronization in the Kuramoto model \cite{arenas06}. The size-turmoil relationship is slightly disturbed by dyads more often facing a larger opponent group and being therefore less likely to fight collectively than triads. If there was one party that dominated, it was always the larger group, with only one exception, in line with earlier models of warfare when controlling for weapons \cite{lanchester56}. 

In our data, bursts developed in 13 (37\%) of the 35 smallest groups (dyads and triads) and in 11 (46\%) of the 24 larger groups. In bursts, the correlation\footnote{The predicted relation is $T_c \approx n^{1/2}$ \cite{bruggeman20a}, but due to the small data and unexplained variance therein, a curve does not fit better than a straight line; hence, the reported correlation.}   
between focal group size and opponents' turmoil is 0.53. 

\begin{figure}[!ht]
\begin{center}
\includegraphics[width=.55\textwidth]{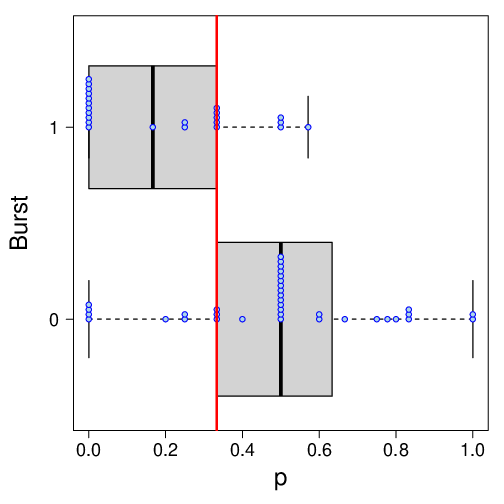} 
\end{center}
\caption{The proportion of non-fighters, and the predicted critical level (vertical line) that separates the boxplots of bursts (1) and non-bursts (0). The data points are our 59 groups.}
\label{fig:result}
\end{figure}

The proportions of non-fighters in groups with bursts (mean~= 0.19; s.d.~= 0.21) and groups without bursts (mean = 0.49; s.d.~= 0.26) are box-plotted in Fig.~\ref{fig:result} (Welch test $t = 4.925; P = 4.119$ x $10^{-6}$; df = 54.41; the ROC-curve has AUC = 0.774).   
The predicted critical threshold (vertical line in the figure) separates the two boxplots, but one might question how robust this result is, given some error in coding. There is a risk of missing a violent act (that can be executed extremely rapidly) or mislabeling a fast movement as violence, thus mixing up fighters and non-fighters, and a risk of mis-assigning individuals to groups. Even if we suppose there is a 10\% chance of each of these errors (which corresponds to Krippendorff's $\alpha = 0.75$; see inter-coder reliability in the Supplementary Material), the result is barely weakened (Welch test $t = 4.010; P = 0.002$, averaged over 1000 simulation runs). Hence, the distinction between bursts and non-bursts is fairly robust. 

The empirical distinction between bursts and non-bursts is not perfect, and Fig.~\ref{fig:result} shows that 14 groups out of 59 (24\%) are categorized incorrectly. Most of the errors are due to local circumstances, not a flawed Ising model.  
In 11 of these 14 groups, the time between the first violent act and last participant joining in was above our two-second interval. Hence, we did not count these collective actions as bursts even though the level of cooperation was higher than expected in non-bursts (Fig~\ref{fig:burst}). One reason for the slow starts could be turmoil lingering at its critical level. To sort this out, we would need a better scale of turmoil than the cumulative scale we currently have. Furthermore, in our 11 slow start cases, local circumstances appear to have overridden the Ising dynamics. When re-examining the cases with time lags beyond two seconds, we noticed various causes of delay: (1) group members were in a car and it took time to get out; (2) a focal group member stood at a distance from the fight on a slippery floor next to a pool; (3) a focal group member was beaten and seemed intoxicated by alcohol, hence responded slowly; (4) a focal group member was first pushed over and it took him time to get on his feet again; (5) opponents moved around so fast that it took time to hit one; (6) an opponent fell into thick bushes such that only one focal group member at a time could get to him; (7) group members were sitting in a train and needed time to get on their feet; (8) group members were constrained in a small room where they searched for space to lash out; and, (9) group members were hindered by the mess of two opponents fighting on a collapsed table full of food. One can hardly blame the model for not incorporating these local circumstances that caused delays, and perhaps our two seconds limit should be applied with more lenience in such cases. Finally, two delays were caused by ambiguous behavior: (10) someone tried to de-escalate, was hit, and then attacked, and (11) someone tried to pull back his fighting friend (de-escalation) but almost simultaneously attacked.

In four cases, non-bursts were predicted whereas bursts occurred.
Although for parsimony, we applied the same $S_0$ (and $\Delta$) for all conditional cooperators, some combatants may actually have had a higher value than the average, $S_{0,i} > \overline{S_0}$, which could explain these cases. 

\subsection{Alternative explanation}
A plausible alternative explanation for the onset of collective action is synchrony of motion \cite{mcneill95}, which yields a feeling of oneness among group members \cite{fischer13} and a stronger willingness to take risks for one's group mates \cite{swann12}. 

We measured the synchronization of focal group members pairwise if they engaged in the following behaviors simultaneously (i.e., overlapping intervals on the timeline): approaching members of the opponent group at the same pace (normal, energetic, or running), distancing from the opponent group at normal walking speed (we consider simultaneous distancing at energetic or running pace uncontrolled attempts to escape rather than forms of synchrony), or simultaneously engaging in aggressing. To calculate the level of synchrony, we noted the duration of each instance of synchrony from the beginning of the clip prior to the moment of maximum participation in violence, and we multiplied it by the proportion of focal group ties involved (i.e., normalized with respect to the maximum number of ties in the group).

Although in 21 out of 23 bursts, some degree of synchronization (10.8 on average) preceded collective violence, there were 18 cases in which synchronization (9.6 on average) was not followed by collective violence. In several of the latter cases, synchrony turned out to be a deceptive performance composed of blustering and aggrandizing without commitment to fighting. This does not imply that synchronization is unimportant (it probably is for solidarity; \cite{collins08}), but it does not predict collective violence. 
 
\section{Discussion and Conclusion} 
The simple Ising model is a century old \cite{brush67} and has been applied to a wide range of problems \cite{stein13,macy24}, to which we add the dilemma of collective action. It explains cooperation parsimoniously, based on agitating stimuli without recourse to strong rationality, initiative takers, norms, feedback through selective incentives, or reliable information passing through the network resulting in reputations. 
The Ising model elucidates the temporal unfolding of violence by predicting a critical threshold of non-fighters that distinguishes a burst of collective action from a slow start. This pattern is largely supported by the data. The model also explains why violent groups are often small or are small subgroups of larger groups despite greater risk. Small (sub)groups have a lower critical threshold of turmoil, and in a confrontation with opponents, lower levels are reached earlier.  We also investigated whether synchronous action precedes violence, but we found that synchronization precedes both collective and solitary violence, and cannot predict either of these outcomes. However, synchronization may still be important to increase solidarity \cite{gelfand20,durkheim12}. 

This study has several limitations. Because we did not select videos without violence, we cannot be certain that violence is caused by turmoil.  Moreover, developing a proper scale of turmoil or agitation for field studies is exceedingly difficult, due to variations across domains (e.g., street violence versus grievances about politics), and different effect sizes of different stimuli (e.g., killed family members versus a push that might be quickly forgotten). On top of these challenges, our measurements underestimated turmoil because the videos are left-truncated, and our data depend on camera angle and vision width that inevitably exclude instances of turmoil.  Measuring turmoil or agitation is only possible in exceptional situations, for example when protest movements produce it by themselves through accelerated posting of online messages. In these particular cases, the inter-posting time intervals follow a certain pattern (Moore's law), which makes it possible to predict $T_c$ ahead of large street demonstrations \cite{johnson16}. In lab experiments,  ethically responsible intergroup conflicts could be examined experimentally, benefiting from controlled circumstances and more accurate measurement of agitation. This can be achieved when subjects engage in a conflict in virtual reality. Then, causality of turmoil and the critical level of non-fighters can be assessed more precisely. 

Furthermore, there might be error in our coding. If the truth were perfectly known and the coding corrected accordingly, the two boxplots in Fig.~\ref{fig:result} might shift to a small degree, but, given our robustness check, it is unlikely that this shift would affect our main result. 

Another limitation is in the Ising model itself. Despite predicting the threshold of non-fighters fairly well, several groups ended up in the non-burst category due to local circumstances (ignored by the model) that caused delays beyond our two second limit. Furthermore, the Ising model does not predict the severity of violence.  For future studies, it is important to expand the number and diversity of cases, and to develop better software than is currently available to automatically code videos.  Finally, when individuals find themselves more often in similar situations, they will learn, which is easier in smaller groups where they have a larger influence on their payoff \cite{burton21}. Some will change their strategy, and turn into non-fighters \cite{andreozzi20} who try to exploit other group members and maximize their individual payoff instead of maximizing the group's payoff. For the model, this would require individuals updating their decision rules at subsequent Monte Carlo steps.  

In this first empirical application, we showed that the Ising model can explain the unfolding of violence, and in all likelihood, more discoveries lay ahead. Extensions to norms (as external field $-h N_m $, with $N_m$ norm enforcers) and noise in actors' information about others' behavior (i.e., reputations) have been explored in simulations \cite{bruggeman24}, but in this empirical study, we had no data about norms and reputations, and left them out of our model. 
In all likelihood, it can be applied to protests \cite{gonzalez11} and revolts, which break out more often if (rumors say that) a government or its police are weakened \cite{skocpol79,tufekci17}, analogous to vulnerable individuals in street fights. In protests, where durations are usually counted in days instead of seconds (as our data), it might be easier to distinguish bursts from slow starts than in our small groups, where the distinction sometimes looked gradual. The model also seems applicable to bystanders collectively helping victims under uncertainty \cite{philpot20}, and to lynchings \cite{beck90}. For example, Nussio \cite{nussio24} argued that lynchings in Mexico can be explained through solidarity combined with peer pressure in a network transmitting reputations meshed with local norms. This argument is perfectly consistent with an expanded Ising model, but it lacks the agitation necessary for cooperation to start without centralized organization, as it did empirically. We would argue that the agitation was due to the violation of social norms, of which the lynching victims were accused (e.g., child theft), augmented by the gossip transmitted through the network. In other countries, putative norm violations that agitated crowds and lead to lynchings were rape \cite{brundage97} and blasphemy \cite{asif23,rumi21}. The Ising model might even be applied to other species, for example herd bulls defending group members against attacking lions, or quorum sensing bacteria \cite{diggle07}. 

Taking a broader evolutionary 
perspective on the role of random noise, multiple studies have found that it can solve coordination problems, such as collective responses to opponents' threats, by shaking a group loose from its suboptimal (e.g., non-cooperative) state \cite{shirado17}, but it can also disturb an optimal state. In the Ising model, both can happen, depending on the amount of turmoil: when in a group, turmoil approaches $T_c$ from below, random noise in decision making facilitates cooperation, but at high levels, too much noise entails confusion and error.    

\subsection*{Data availability}
All coded video data, the R script used to produce plots from the data (Supplementary Material), and a Fortran script for simulations of the Ising model are available at \texttt{https://osf.io/f25nq/}

\subsection*{Authors' contributions}
JB made the asymmetric Ising model, and wrote the paper and all software. DW collected, interpreted, and (after JB ploted the graphs such as Fig.~S4) analyzed the data. BM did the mean field analysis. JB and DW wrote the Supplementary Material.

\newpage
\begin{center}
\section*{---Supplementary Material---}
\end{center}

\section*{Data and compilation of data set}
In contrast to prior studies of violence and deescalatory action based on CCTV footage \cite{levine11,liebst19,philpot20} most of our videos were captured on mobile phones, which tend to provide better picture clarity, detail and sound. Whereas CCTV footage is most often shot from an elevated and fixed camera angle, people who record violent incidents on their mobile phones tend to follow the action, thereby allowing us to observe bodily positioning and movements in more detail \cite{luff12}.

One concern about phone-recorded clips is that recorders start filming when the antagonism is already ongoing; our data are thus left-truncated to an unknown degree. We discarded footage that immediately started with physical violence. Another concern is that uploaded videos might be biased toward more spectacular cases. However, our sample shows variety in the forms and severity of violence, ranging from groups who engage in frenzied collective violence to incidents that involve just a few slaps. A third concern is whether the recording influenced the behavior of the assailants. Several videos showed more than one person recording the incident. A host of research indicates that people do not change their behavior substantially in the presence of a camera and mainly focus on what they are doing (reviewed by \cite{jones12}). This is probably even more the case in antagonistic situations, and because young people are used to being filmed by phones. Additionally, when people do pay attention to the camera, which happened in just one of our clips, it is visible in the recording and can be incorporated in the analysis. (In our case, this concerned behavior coded as turmoil.) 

We used the following criteria to compile our dataset. First, we only included incidents in which at least one of the antagonistic parties comprised at least two members. Second, we only selected clips in which violence could be observed from the (or a) beginning until the end. Third, we focused on fights that seemed to have occurred spontaneously, excluding prearranged fights on the basis of any of the following criteria: the fighters and bystanders all had approximately the same young age (e.g., school yard fights); the antagonists were wearing protective clothing; or a referee was present. Regarding gender, the overwhelming majority of the incidents were between males, with only 8 cases involving both females and males. We excluded female-only fights because they often seemed being arranged (back yard fights with male referees).

Along these criteria, we obtained 42 videos from websites such as YouTube, LiveLeak, and WorldStarHipHop, using search terms with the English keywords ``brawl,'' ``street fight,'' and ``assault.'' This sample appears to be random with respect to temporal unfolding and (sub)group size. 36 clips are from English-speaking countries (mainly the US and the UK, with one from Canada and one from India); 5 of the remaining clips are from the Netherlands, and 1 is from Colombia. We did not observe differences in relevant behavior related to the location of the recording. 

The shortest clip lasted 30 sec.~and the longest nearly 5 minutes (mean 101 sec.; s.d.~59 sec.). Out of a potential 2 x 42 groups, where the opponents in one analysis become the focal group in the next, 25 groups fought with a single individual rather than a group, which leaves $\mathcal{N}$=59 groups to examine. Most groups were small, $2 \leq n < 10$ (mean 3.6), but one had 14 members (of which 6 became violent). The smaller ones were simulated as fully connected networks wherein everyone could see one another unless there were obstacles or deescalators obstructing visual contact. Obstruction was simulated by removing $m$ ties. 

\section*{Coding}
We used Noldus Observer XT 14 software to code the behaviors of individuals, resulting in start times and durations of behaviors per individual throughout the video clip. The software distinguishes state events of relatively longer duration (e.g., holding a person) from point events that are brief single acts (e.g., a punch). When an act of violence took a more extended duration (e.g., holding or dragging), we coded the entire duration of the act, thus creating a state event. When a series of violent point events appeared less than two seconds after one another, we connected them, thus creating a state event.

To enhance the precision of our coding, we coded observable actions (such as kicking, punching, pointing at opponent) and later lumped them together to generate the main codes: violence, turmoil, and deescalation. We coded the behavior of each actor in multiple rounds, coding only one type of action per round. The video material was played at half speed and repeated many times, which is important because violence proceeds too fast to see what is going on in real time. The first 20 clips were coded by one of the authors and two assistants. We discussed the coding when we disagreed and continued the procedure by multiple coders until agreement was reached. At this point, the remainder of the clips were coded by one (meanwhile) experienced assistant researcher and were later checked by another. The remaining disagreements or mistakes were discussed, and the coding was adjusted accordingly.

To identify whether individuals belonged to a focal, opponent, or third-party group, we observed whether they remained close (at touching distance) to each other as they moved through space, whether they called each other by their names, whether they touched each other (for instance, by tapping shoulders), or whether they shared a vehicle or other object (for instance, a bag or skateboard). One of the authors and two assistants discussed each case in which we observed one or more of these behaviors to determine whether the actors were part of the same group.  To assess the reliability of our coding, we asked three students who were unaware of our theoretical expectations and preliminary results to code group memberships, maximum number of participants involved in violence, and the occurrence of bursts in 37 videos, containing 54 groups and 20 solitary individuals (according to our own coding). Notice that just like biologists observing primates, much experience is required to become an accurate observer, which these students did not have. Krippendorff’s alphas for group membership, fighters, and bursts were .87, .68, and .75 respectively, or .77 on average; in the main text we reported that an error percentage of 10\% results in .75. We also calculated whether the distinction between bursts and fizzles in students’ coding followed the predicted critical proportion of non-fighters. We arrived at the number of non-fighters (which the students did not code) by deducing the maximum number of participants in violence from the group size.  Out of 49 groups (5 cases missing or coded as individual rather than group, as we did) students identified 29 bursts and 20 fizzles (we required that at least 2 out of 3 students agreed on the observation of a burst to denote it as such). The proportion of defectors was .18 (sd $= .03$) in bursts and .33 (sd $= .04$) in fizzles (Welch test 8.3, df 35.6, $P=.007$), with bursts clearly below and fizzles right at our critical threshold. 

\section*{Composition of indicators}
\textit{Violence}. We coded violence when members of the focal group used force against another's person's body (by punching, slapping, kicking, hitting, stomping) and/or when actors moved or held another person's body forcefully (by pushing, shoving, dragging, wrestling, holding, etc.). To enhance the reliability of our coding, we also coded whether actors used their hands, legs, or other body parts (e.g., elbow, head) and whether weapons were used, including the type of weapon, excluding guns (i.e., knives, sticks, tasers and improvised weapons, such as a skateboard or bottle, etc.). We define collective violence as time intervals wherein two or more members of the focal group used violence simultaneously, either as overlapping state events or, in fewer cases, as simultaneously occurring point events. The maximum participation in collective violence was determined by taking the largest number of focal group members engaging in violent action, divided by the total number of focal group members. 

In the videos, it was not possible to distinguish leaders from initiative takers, but we noticed individuals who started violence on their own. In simulations, a leader/initiative taker $i$ can be incorporated by a larger, individual $S_{0,i}$ value; the result is collective action at lower $T$.

\textit{Non-fighters} do not cooperate when $T \ge T_c$ or most network neighbors cooperate. First, we labeled as non-fighters, the focal group members who took deescalatory action towards others. 
Their behavior was coded as follows \cite{levine11}: open-handed gestures in the direction of other individuals; waving arms to stop or dampen in the direction of others; touching or patting; guiding a person away; pulling people apart; and putting one's body in between opponents. 
Whether deescalators' interventions were effective or not, they were at least temporarily unavailable to participate in violence themselves, and their behavior signaled noncooperation to their fellow group members. Second, we noted group members who were effectively stopped from using violence for at least five seconds by other focal group members, opponents, or third-party members (e.g., bystanders). Third, group members remaining passive when others fought. Fourth, group members unable to participate in violence due to spatial constraints for at least five seconds, for instance, due to cars blocking their way or because they were not close enough to the action. Fifth, group members unable to participate for at least five seconds because they were wounded or had fallen to the ground. We took all five reasons for and causes of defection into account for the proportion of non-fighters before and during the first instance of violence by the focal group.

\textit{Turmoil}. We subsumed the following behaviors of members of the opponent group under the heading of turmoil for the focal group. First, we coded approaching when opponents moved closer to focal group members. We also indicated whether actors moved energetically (jumping/skipping) and whether they ran. Second, aggressive behaviors when actors in the opponents' group moved body parts in a way that signaled provocation, hostility and/or a readiness to attack. We used the following modifiers (sub-codes) to specify aggressing behavior: fighting gestures; pulling off clothing (jackets or vests); pulling up pants; pointing at opponents; provocative gesturing with fingers or hands (as an invitation to engage); bending forward (head and/or chest bending toward opponent); encroaching (invading opponents' personal space through using or damaging objects belonging to them); and teasing (invading opponents' body space in a teasing/ridiculing way that does not inflict bodily harm, such as lightly hitting, stroking, or patting the opponent's body). Third, involuntary behavior that increases the vulnerability of opponents, because it tends to agitate (signaling opportunity) and provoke violence. This concerned stumbling and falling to the ground. Finally, violence committed by members of the opponent group.
Because we do not know the psychological effect sizes of the various kinds of turmoil, we could not construct a ratio level scale. Instead, we calculated the total level of turmoil from the beginning of the clip until a focal group's maximum participation in violence by noting the duration of each instance of turmoil, and multiplying it by the number of opponents involved, i.e., the surface area under the turmoil curve in Fig.~S\ref{fig:S4}, S\ref{fig:S7}, and S\ref{fig:S8}.  

\textit{Bursts}. We identified bursts when, at the first moment of collective violence, two conditions are fulfilled: (1) at least half of the focal group members joined the fight, or both actors did in a dyad, and (2) cooperators started fighting in less than 2 seconds (with a 5\% margin) after the first. Condition (1) follows from simulations of small groups; see Table~S\ref{table:defect}. Note that in simulations of much larger networks ($n \geq 1000$), bursts can occur wherein less than half of a group participates. In some videos, we observed multiple bursts; in these cases, we only considered the first burst to facilitate comparison of cases.

\section*{Examples from videos}
\begin{figure}[ht!]
\begin{center}
\includegraphics[width=.8\textwidth]{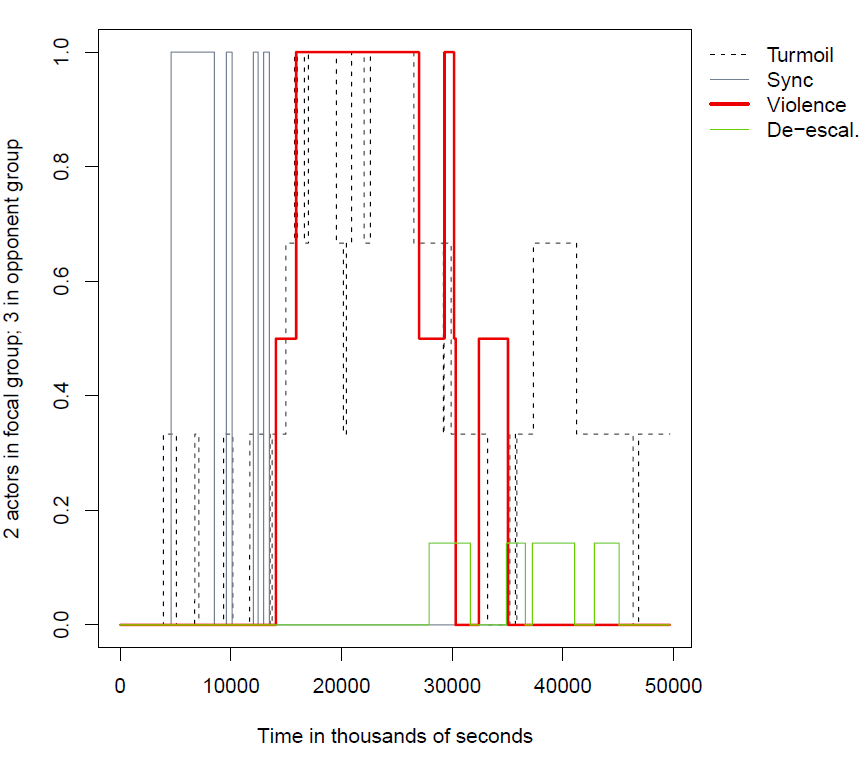} 
\end{center}
\caption{ Graph of clip 26 with a burst of collective violence by a dyad. } 
\label{fig:S4}
\end{figure}

\begin{figure}[ht!]
\begin{center}
\includegraphics[width=.8\textwidth]{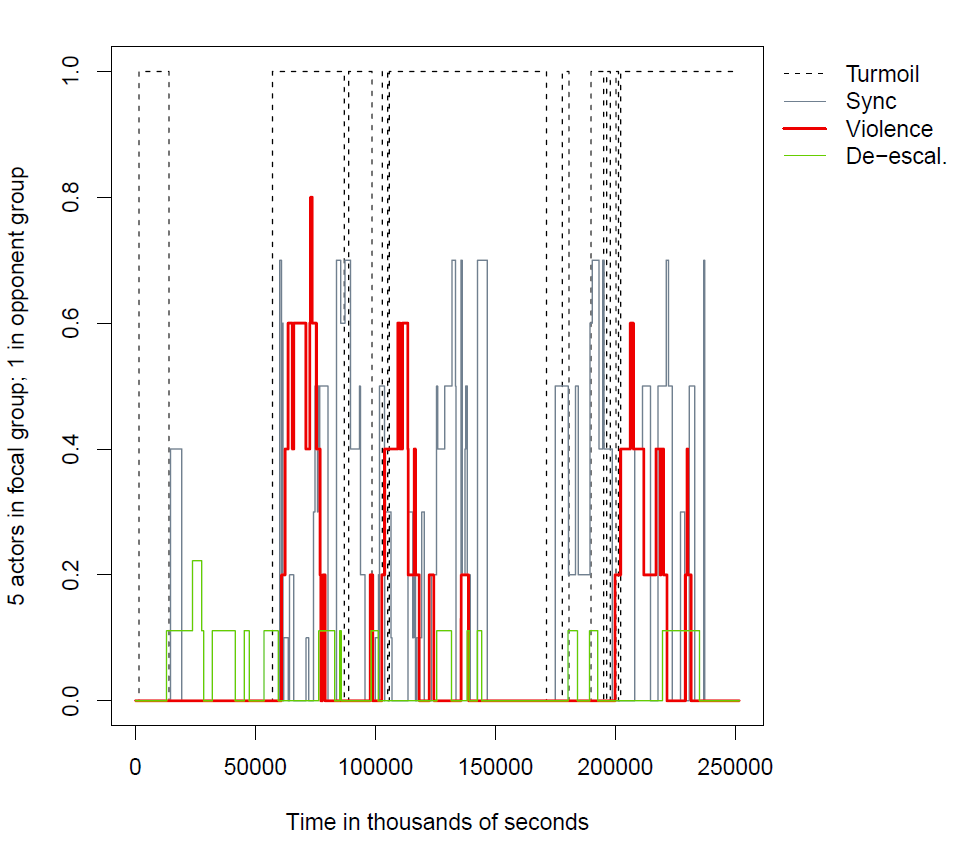} 
\end{center}
\caption{Graph of clip 86. Outbreaks of collective violence in a group of five. Because the time between the first and last participant's commencement of violence exceeds 2 seconds, these outbreaks are not classified as bursts. } 
\label{fig:S7}
\end{figure}

Graphs of behavior on the timeline enable to analyze and compare the coded videos. Each graph shows the intervals and their durations of (collective) violence, turmoil, synchrony and deescalation (see Fig.~S\ref{fig:S4}, S\ref{fig:S7}, S\ref{fig:S8}).

\begin{figure}[ht!]
\begin{center}
\includegraphics[width=.8\textwidth]{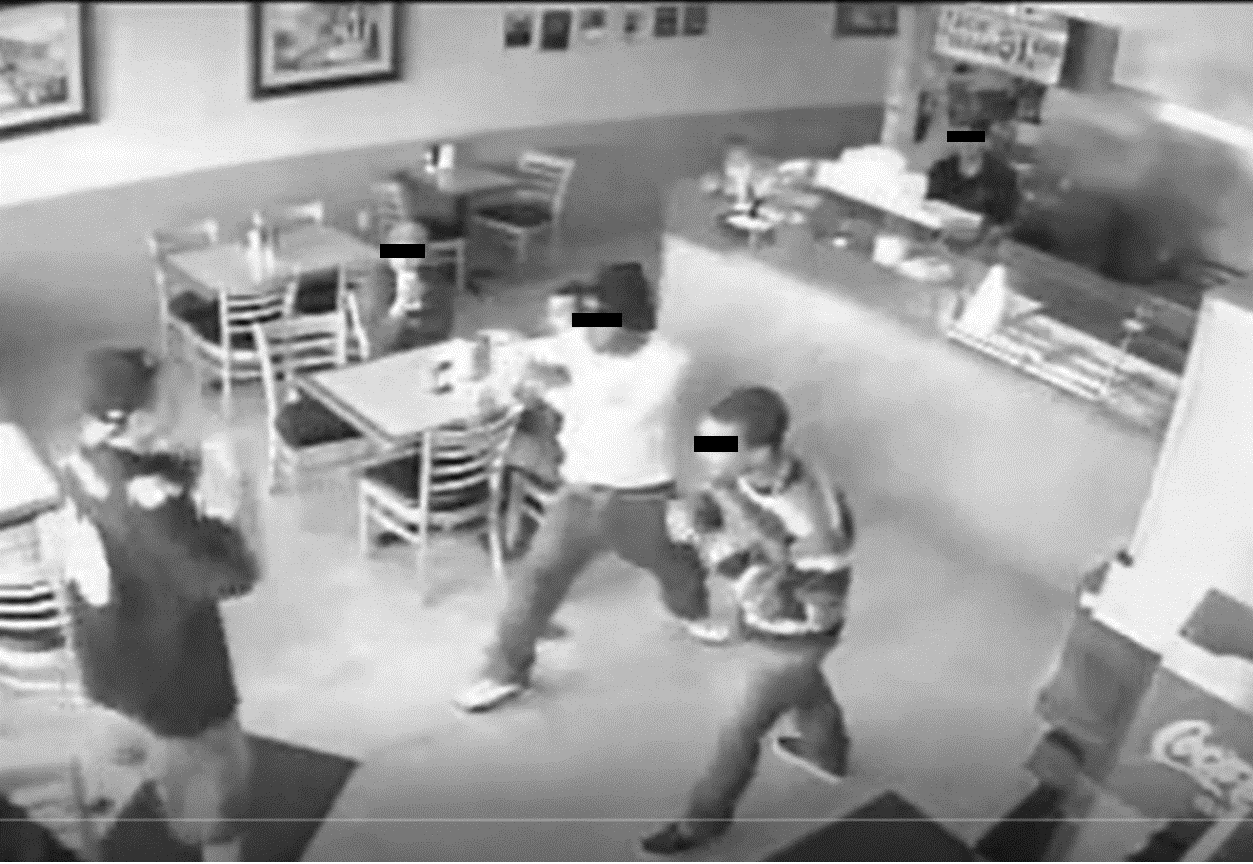} 
\end{center}
\caption{Still taken from clip 26 at 0:06 seconds.} 
\label{fig:S1}
\end{figure}

\begin{figure}[ht!]
\begin{center}
\includegraphics[width=.8\textwidth]{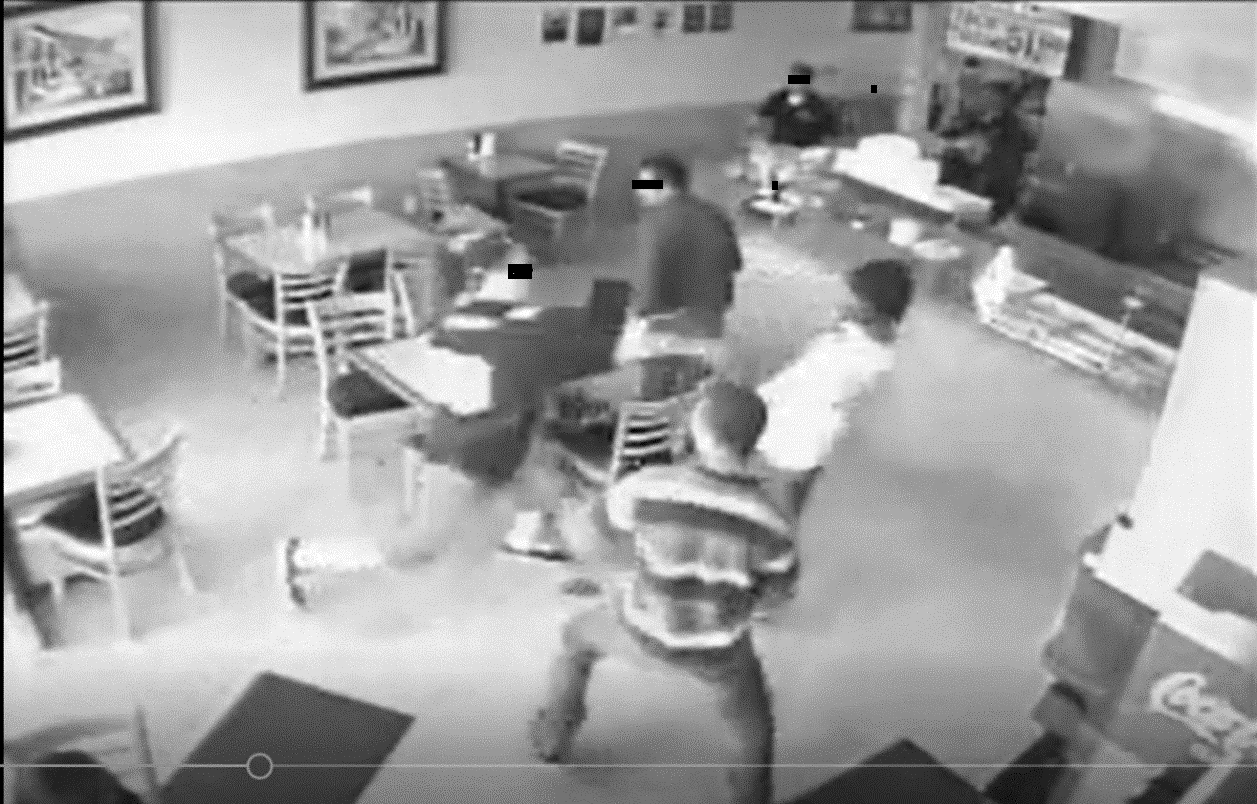} 
\end{center}
\caption{Still taken from clip 26 at 0:13 seconds. } 
\label{fig:S2}
\end{figure}

\begin{figure}[ht!]
\begin{center}
\includegraphics[width=.8\textwidth]{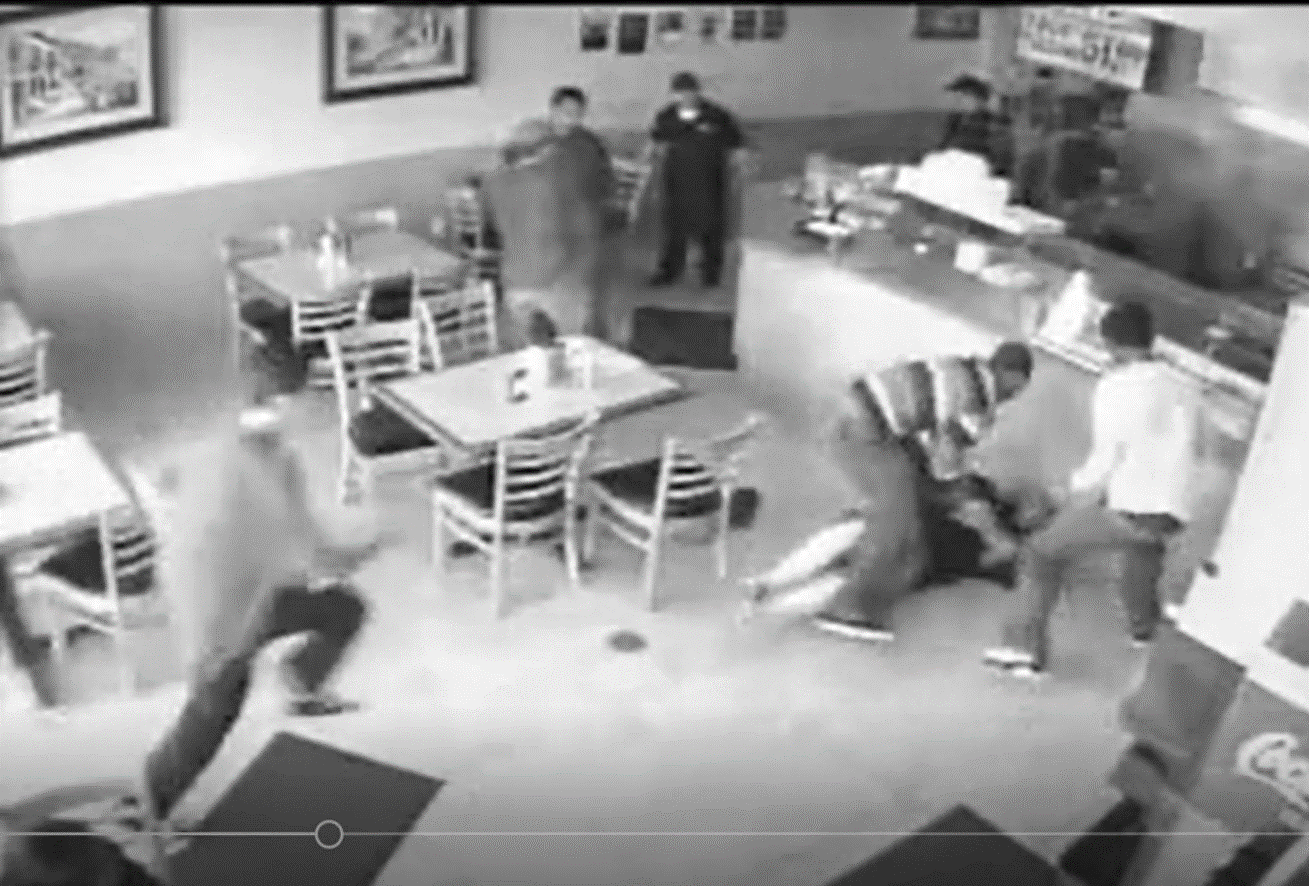} 
\end{center}
\caption{ Still taken from clip 26 at 0:15 seconds. } 
\label{fig:S3}
\end{figure}

To illustrate, we discuss how coded video data was turned into a graph for clip 26. Fig.~S\ref{fig:S1}-\ref{fig:S3} feature screenshots of the clip, and Fig.~S\ref{fig:S4} displays the graph. This clip features a confrontation of a dyad (focal group) with a triad (opponents) in a restaurant. Violence started at approximately 14 seconds into the video and quickly became collective until approximately 28 seconds. Four instances of turmoil preceded the violence, starting with one of the members of the opponent group gesturing aggressively and walking around in a jumpy, energetic way. The members of the focal dyad walked synchronously toward and away from the opponent in three shorter and one relatively longer stretch of time. Fig.~S\ref{fig:S1} captures a moment of synchrony; the members of the dyad were making fighting gestures and stood aligned, each stretching out one leg with their feet pointing toward the opponent, and the other leg standing backward. Fig.~S\ref{fig:S2} shows the moment when the dyad's level of agitation reached a critical level; when one of the focal actors dodged backward, avoiding being hit, his companion was about to deliver a blow to the head of the opponent, which made the latter fall down. Fig.~S\ref{fig:S3} displays the second member of the opponents' triad (soon joined by the third) rushing into the restaurant toward the focal dyad, near his fallen companion. At this point, both groups were engaged in collective violence. Fig.~S\ref{fig:S2} also shows bystanders (who can be seen sitting at a table in Fig.~S\ref{fig:S1}) and employees standing near the counter. At 28 seconds, an employee took deescalatory action for the first time. At that point, the collective violence had stopped already and erupted again for a shorter moment approximately 2 seconds later. The clip ends when two triad members carry their fallen group member outside. By then, the focal dyad had already left the scene. 

\begin{figure}[ht!]
\begin{center}
\includegraphics[width=.8\textwidth]{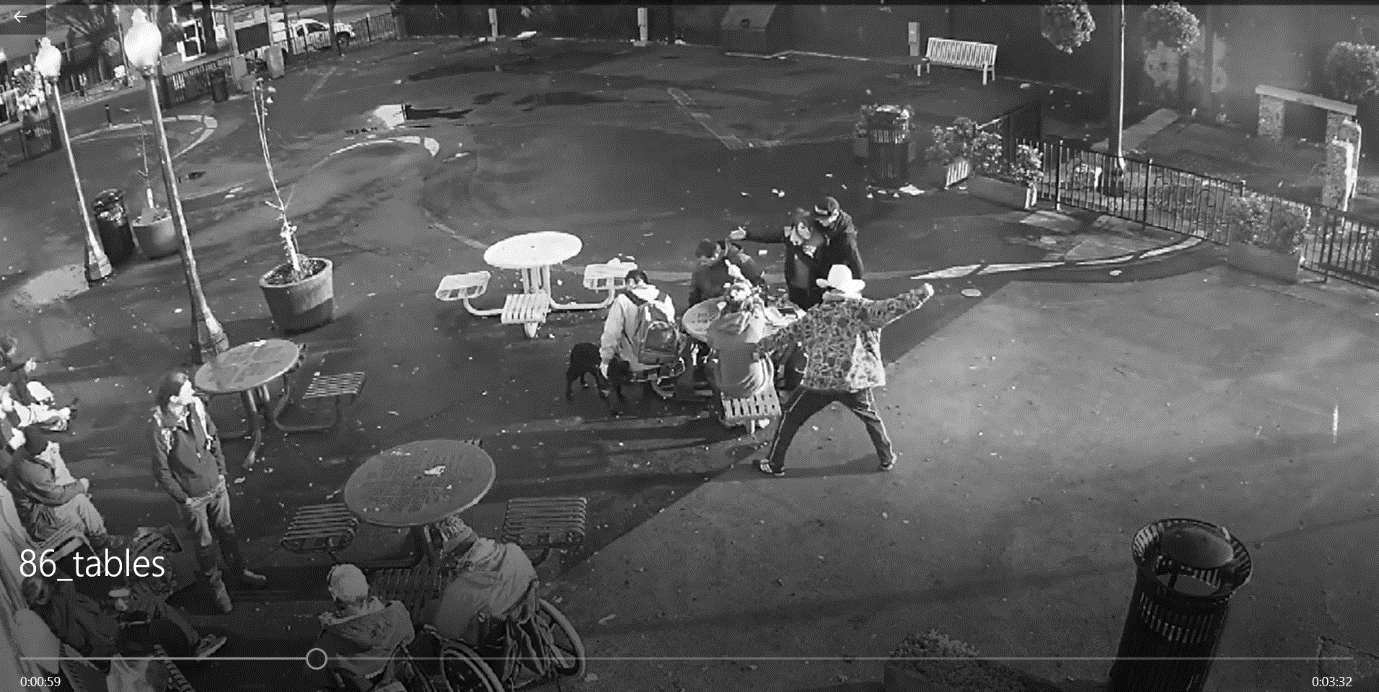} 
\end{center}
\caption{ Still taken from clip 86 at 0:59 seconds. } 
\label{fig:S5}
\end{figure}

\begin{figure}[ht!]
\begin{center}
\includegraphics[width=.8\textwidth]{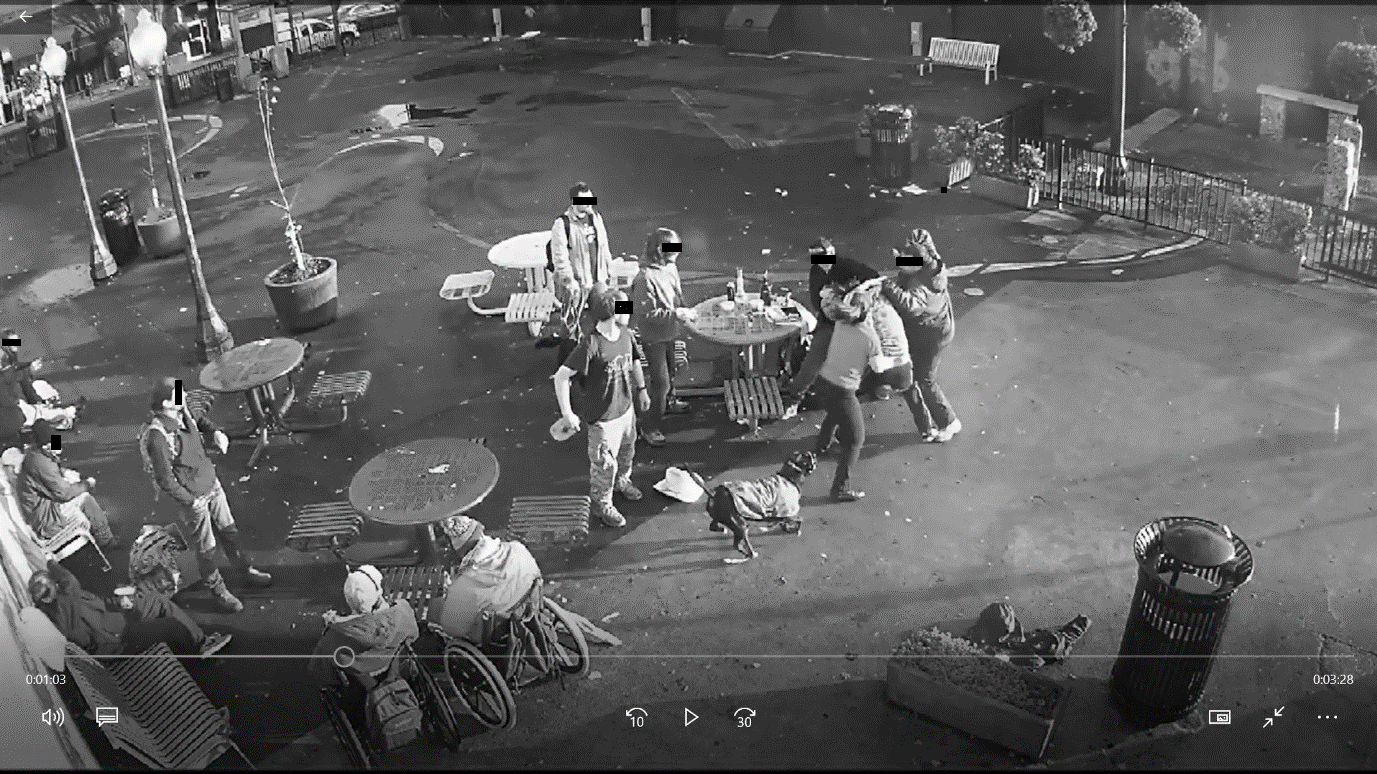} 
\end{center}
\caption{ Still taken from clip 86 at 1:03 seconds. } 
\label{fig:S6}
\end{figure}

\vspace{0.7cm}
Fig.~S\ref{fig:S5}-S\ref{fig:S7} provide another illustration. Fig.~S\ref{fig:S7} is the graph of clip 86; it shows three outbreaks of collective violence in a group of five members who faced a single opponent. In the first instance of collective violence (not a burst by our 2 seconds criterion), starting at approximately 60 seconds into the clip and lasting for approximately 30 seconds, four members participated. Three of them immediately followed each other, and the fourth joined after approximately 2 seconds. Prior to the outbreak of violence, the opponent agitated focal group members by walking back and forth and by posturing aggressively. Fig.~S\ref{fig:S5} captures the trigger moment of agitation; i.e., while one member of the focal group was engaged in aggressive posturing and a group member tried to deescalate, the opponent was about to strike another focal group member. Fig.~S\ref{fig:S6} shows the situation 4 seconds later, when three members of the focal group were attacking.

\section*{Analysis of coded and plotted videos}
\begin{figure}[ht!]
\begin{center}
\includegraphics[width=.9\textwidth]{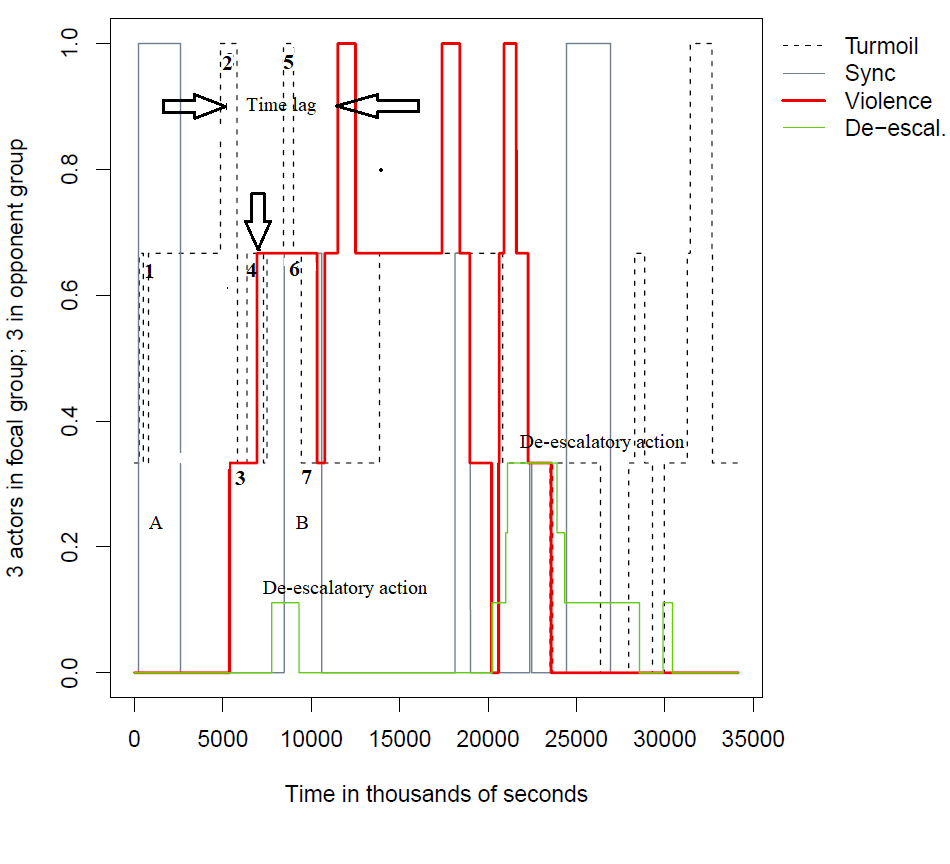} 
\end{center}
\caption{Graph of clip 67; opponent group. The timeline displays the start of collective violent action (downward arrow), instances of turmoil (1-7), synchrony (A,B), the time lag between the first and last group member joining the collective violence (horizontal arrows), and deescalatory action.  } 
\label{fig:S8}
\end{figure}

We used graphs of coded videos in our analysis, illustrated in Fig.~S\ref{fig:S8}, where the opponent group in clip 67 is plotted. The horizontal axis is the timeline in thousands of seconds, the vertical axis is the proportion of group members involved (for turmoil and violence), and the proportion of intragroup ties involved out of the maximum possible number of ties (for synchrony of action). The first moment of collective violence, denoted by the downward arrow, involves two focal group members. There is, however, a short interruption of collective violent action. Therefore, we did not code this occurrence of collective violence as a burst. The time gap between the first participant and maximum participation in collective violence, denoted by the horizontal arrows, takes over 6 seconds.

In the figure, seven instances of turmoil occur (numbered 1-7) prior to the moment of maximum participation in violence: (1) a short interval that, after a brief interruption, continues for 4 seconds and involves 67\% of the opponent group (2 members); (2) involves all three members of the opponent group and lasts nearly 1 second; (3) lasts less than 1 second and involves 33\% of the opponent group; (4) turmoil continues, with an intermittent short dip, for over 1 second with 67\% involvement; (5) a peak at 100\% involvement that lasts less than 1 second; (6) a short drop to 67\% involvement that takes approximately 1 second; (7) the last turmoil before maximal violence. We calculated the level of turmoil by multiplying the time length of each of these intervals by the number of actors involved.

Intervals of synchrony in Fig.~S\ref{fig:S8} are indicated by A and B. A lasted 2.4 seconds and involved all ties in the group; B involved 67\% of the ties for approximately 2 seconds prior to maximum participation in violence. Finally, the graph shows intervals of deescalatory action.

\newpage

\section*{Overview of cases and descriptive statistics}
Our dataset contains information about the behavior of 406 individuals in 42 violent incidents. The 59 groups and their key characteristics are listed in Table~S\ref{table:long}.
After focal and opponent groups have been analyzed as listed, they flip roles with respect to cooperation and turmoil production in a second batch of the analysis, except when an opponent is solitary instead of a group. 

\begin{footnotesize}
\begin{longtable}{rlrrrrrrrr}
  \hline
  & group & $n$ & c.v. & $N_C$ & lag & burst & $T$ & Sync & $N_D$ \\ 
  \hline
1 & 1\_beachfight\_focal &   3 &   1 &   2 & 0.00 &   1 & 0.97 & 19.39 & 1 \\ 
  2 & 3\_asianeateries\_focal &   5 &   0 &   1 &  &   0 & 72.41 & 18.10 & 4 \\ 
  3 & 3\_asianeateries\_oppon. &   2 &   0 &   1 &  &   0 & 93.37 & 16.20 & 1\\ 
  4 & 6\_rikshadrivers\_focal &   6 &   1 &   4 & 1.93 &   1 & 7.24 & 5.33 & 2 \\ 
  5 & 7\_baldies\_focal &   4 &   1 &   3 & 0.92 &   1 & 6.42 & 0.94 & 1 \\ 
  6 & 8\_queensday\_focal &  14 &   1 &   6 & 1.95 &   1 & 45.81 & 303.42 & 8 \\ 
  7 & 9\_gasstation\_focal &   2 &   0 &   1 &  &   0 & 14.49 & 16.30 & 1 \\ 
  8 & 10\_redpickup\_focal &   3 &   1 &   3 & 0.93 &   1 & 9.34 & 78.75 & 0 \\ 
  9 & 26\_insideeatery\_focal &   3 &   1 &   3 & 1.21 &   1 & 14.41 & 0.20 & 0 \\ 
  10 & 26\_inseeatery\_oppon. &   2 &   1 &   2 & 1.77 &   1 & 7.49 & 5.32 & 0 \\ 
  11 & 27\_dallasairport\_focal &   6 &   1 &   5 & 0.00 &   1 & 24.99 & 3.53 & 1 \\ 
  12 & 32\_shovedbycar\_focal &   3 &   1 &   2 & 3.66 &   0 & 3.66 & 0.00 & 1 \\ 
  13 & 41\_kababshop\_focal &   2 &   0 &   1 &  &   0 & 42.12 & 0.00 & 1 \\ 
  14 & 41\_kababshop\_oppon. &   2 &   1 &   2 & 0.84 &   1 & 5.90 & 0.42 & 0 \\ 
  15 & 45\_onbusystreet\_focal &   2 &   1 &   2 & 1.48 &   1 & 12.79 & 9.35 & 0 \\ 
  16 & 50\_fitness\_focal &   2 &   1 &   2 & 3.08 &   0 & 4.62 & 1.54 & 0 \\ 
  17 & 52\_blockingskaters\_focal &   2 &   0 &   1 &  &   0 & 20.44 & 10.72 & 1 \\ 
  18 & 52\_blockingskaters\_oppon. &   4 &   0 &   1 &  &   0 & 45.56 & 17.36 & 3 \\ 
  19 & 53\_famousonyoutube\_focal &   2 &   1 &   2 & 20.23 &   0 & 14.45 & 0.00 & 1 \\ 
  20 & 53\_famousonyoutube\_opp. &   2 &   0 &   1 &  &   0 & 51.11 & 0.00 & 1 \\ 
  21 & 54\_hammeronhead\_focal &   4 &   1 &   3 & 1.57 &   1 & 9.17 & 4.19 & 1 \\ 
  22 & 54\_hammeronhead\_oppon. &   2 &   1 &   2 & 1.57 &   1 & 31.43 & 5.24 & 0 \\ 
  23 & 60\_gangattack\_focal &   6 &   0 &   1 &  &   0 & 31.73 & 80.43 & 5 \\ 
  24 & 61\_burgerking\_focal &   2 &   0 &   1 &  &   0 & 37.88 & 0.00 & 1 \\ 
  25 & 66\_guyvsgang\_focal &   6 &   1 &   6 & 1.81 &   1 & 0.91 & 19.22 & 0 \\ 
  26 & 67\_poolfight\_focal &   3 &   1 &   2 & 0.84 &   1 & 16.46 & 4.36 & 1 \\ 
  27 & 67\_poolfight\_oppon. &   3 &   1 &   3 & 6.29 &   0 & 19.87 & 2.49 & 1 \\ 
  28 & 84\_brokenwindow\_focal &   3 &   0 &   1 &  &   0 & 35.36 & 2.89 & 2 \\ 
  29 & 86\_tables\_focal &   5 &   1 &   4 & 2.25 &   0 & 29.23 & 4.50 & 1 \\ 
  30 & 88\_babypowder\_focal &   2 &   1 &   2 & 2.10 &   1 & 5.43 & 4.20 & 0 \\ 
  31 & 92\_didntdoanything\_focal &   2 &   0 &   1 &  &   0 & 46.51 & 8.35 & 1 \\ 
  32 & 93\_groupkicking\_focal &   4 &   1 &   3 & 5.92 &   0 & 8.68 & 10.02 & 1 \\ 
  33 & 95\_apologise\_focal &   9 &   1 &   2 & 1.51 &   0 & 26.42 & 76.30 & 7 \\ 
  34 & 95\_apologise\_oppon. &   2 &   0 &   1 &  &   0 & 133.06 & 2.64 & 1 \\ 
  35 & 96\_gangversusbat\_focal &   6 &   0 &   1 &  &   0 & 25.00 & 67.06 & 5 \\ 
  36 & 97\_alleyfight\_focal &   4 &   0 &   1 &  &   0 & 61.14 & 26.32 & 4 \\ 
  37 & 97\_alleyfight\_oppon. &   4 &   1 &   2 & 0.76 &   1 & 34.39 & 3.83 & 2 \\ 
  38 & 98\_stopmick\_focal &   3 &   1 &   2 & 0.76 &   1 & 20.61 & 17.54 & 1 \\ 
  39 & 98\_stopmick\_oppon. &   2 &   0 &   1 &  &   0 & 68.10 & 10.70 & 1 \\ 
  40 & 101\_fightandrob\_focal &   6 &   1 &   4 & 2.97 &   0 & 22.19 & 21.45 & 2 \\ 
  41 & 101\_fightandrob\_oppon. &   2 &   0 &   1 &  &   0 & 79.09 & 3.10 & 1 \\ 
  42 & 104\_waiting\_focal &   4 &   1 &   4 & 0.73 &   1 & 13.83 & 0.36 & 0 \\ 
  43 & 104\_waiting\_oppon. &   2 &   0 &   1 &  &   0 & 128.82 & 0.36 & 1 \\ 
  44 & 105\_frenziedassault\_focal &   3 &   1 &   2 & 1.21 &   1 & 0.00 & 9.61 & 1 \\ 
  45 & 108\_sidewalkbrawl\_focal &   2 &   0 &   1 &  &   0 & 268.40 & 44.73 & 1 \\ 
  46 & 108\_sidewalkbrawl\_oppon. &   4 &   1 &   2 & 0.00 &   1 & 33.55 & 36.72 & 2 \\ 
  47 & 110\_nietklaar\_focal &   5 &   1 &   3 & 14.45 &   0 & 36.13 & 12.04 & 2 \\ 
  48 & 121\_eightballjacket\_focal &   4 &   1 &   3 & 2.66 &   0 & 8.64 & 1.08 & 1 \\ 
  49 & 124\_leavehimalone\_focal &   3 &   1 &   3 & 4.86 &   0 & 13.36 & 0.00 & 0 \\ 
  50 & 124\_leavehimalone\_oppon. &   2 &   1 &   2 & 0.00 &   1 & 9.09 & 0.00 & 0 \\ 
  51 & 128\_throwingfood\_focal &   2 &   1 &   2 & 1.14 &   1 & 0.57 & 0.00 & 0 \\ 
  52 & 128\_throwingfood\_oppon. &   2 &   1 &   2 & 2.28 &   0 & 6.83 & 0.00 & 0 \\ 
  53 & 129\_awayfrommyfood\_focal &   5 &   1 &   2 & 3.06 &   0 & 14.22 & 4.09 & 3 \\ 
  54 & 129\_awayfrommyfood\_oppon. &   6 &   0 &   1 &  &   0 & 65.37 & 30.00 & 5 \\ 
  55 & 133\_homeboy\_focal &   2 &   1 &   2 & 2.68 &   0 & 0.54 & 1.07 & 0 \\ 
  56 & 134\_down\_focalgroup &   4 &   1 &   2 & 0.89 &   1 & 23.12 & 45.27 & 2 \\ 
  57 & 134\_down\_oppon. &   5 &   1 &   2 & 0.89 &   0 & 40.02 & 9.34 & 3 \\ 
  58 & 138\_cometomyhouse\_focal &   3 &   0 &   0 &  &   0 & 34.02 & 36.51 & 3 \\ 
  59 & 145\_knifefightcolombia\_focal &   2 &   0 &   1 &  &   0 & 17.04 & 20.16 & 1 \\ 
   \hline 
\label{table:long}
\end{longtable}
\end{footnotesize} 
Table~S\ref{table:long}: Overview of the groups. Occurrence of collective violence (c.v.); maximum participation in violence ($N_C$); time lag between the first participant and maximum participation in collective violence; burst; levels of opponents-produced turmoil ($T$) and of synchrony at the first outbreak of c.v.; and, number of non-fighters ($N_D$).
In case 6, two focal group members disappeared from view before the first moment of violence had started; out of the remaining 12, 6 participated in collective violence. In cases 19, 27 and 36, $NC + ND \neq n$ due to participants' role switching. In cases 19 and 27, one of the group members defected first (deescalated; was not present at the start) but then switched to fighting. In case 36, the single fighter fell to the ground.

Table~S\ref{table:overview} provides mean values, percentages, standard deviations, and minimum and maximum values for the indicators used in the analysis. 
\begin{table}[ht]
\begin{footnotesize}
\begin{tabular}{rlrrrr}
  \hline
   & & mean or $n$ (\%) & s.d. & min & max \\ 
  \hline
  1 & Group size & 3.58 & 2.09 & 2 & 14 \\ 
  2 & Dyads & 24 (40.7) & & 0 & 1 \\ 
  3 & Triads & 11 (18.6) & & 0 & 1 \\ 
  4 & Groups $\ge 4$ members & 24 (40.7)& & 0 & 1 \\ 
  5 & Collective violence & 38 (64.4) & & 0 & 1 \\ 
  6 & Burst & 23 (39) & & 0 & 1 \\ 
  7 & Maximum participation in violence & .64 & .28 & 0 & 1 \\ 
  8 & Average time lag in bursts in sec. ($n=23$) & 1.06 & .65 & 0 & 2.10 \\ 
  9 & Average time lag per participant in non-bursts ($n=15$) & 5.12 & 5.29 & 0.89 & 20.23 \\ 
  10 & Turmoil prior to max participation in violence ($n=38$) & 37 (97.4) & & 0 & 1 \\ 
  11 & Level of turmoil prior to max part.~in violence ($n=38$) & 15.34 & 12.21 & 0 & 45.81 \\ 
  12 & Synchrony prior to max part.~in violence ($n=38$) & 32 (84.2) & & 0 & 1 \\ 
  13 & Level of synchrony prior to max part.~in violence ($n=38$) & 18.98 & 50.95 & 0 & 303.42 \\ 
  14 & Proportion non-fighters & .37 & .28 & 0 & 1.0 \\ 
   \hline
\end{tabular}
\end{footnotesize}
\caption{Descriptive statistics. Note: Levels of synchrony prior to maximum violence includes an outlier of 303.42. Without outlier, mean = 11.29; s.d.~= 18.96; max = 78.75. We excluded this outlier in the calculations reported.  }
\label{table:overview}
\end{table}

\section*{Solving the Ising model}
The Ising model can be solved computationally by the Metropolis algorithm (main text).
Here, it is solved analytically by a mean field analysis, first for $p = 0$ and then $p > 0$. Subsequently, deviations of the computational results from the mean field result are pointed out.

\subsection*{Mean field analysis}
For the mean field analysis, we assess the level of cooperation, $N_C/n$, in terms of the order parameter, $M = 1/n \sum_{i=1}^{n} S_i$. Consequently, $N_C/n = (M + D)/(C + D)$.  The mean field assumption can be stated as $S_i = \bar{S_i} = M$. For the moment, the proportion of non-fighters, $p = 0$; see below when $p > 0$.

We start out with the Hamiltonian, $H = -\sum_{i,j} w_{ij}S_iS_j$. We use the mapping  
$\{C, -D\} \rightarrow \{ S_0 + \Delta, S_0 - \Delta \}$ with bias $S_0 = (C-D)/2$ and offset $\Delta = (C+D)/2$ to rewrite the Hamiltonian as 
\begin{equation}
    H = -\sum_{i,j} w_{ij}(S_0+\hat{S_i})(S_0+\hat{S_j}),
\end{equation} 
with $\hat{S_i}$ and $\hat{S_j} \in \{-\Delta, \Delta\}$, and taking into account the 
row-normalization of the adjacency matrix ($\sum_{j}^n w_{ij} = 1$).

To calculate the Boltzmann probabilities of a single spin (or an individual's probabilities to cooperate or defect), we define the pertaining Hamiltonian 

\begin{align}
H_i &= -\sum_j w_{ij}(S_0+\hat{S_i})(S_0+\hat{S_j})\\
H_i &= -\sum_j w_{ij}(S_0+\hat{S_i})M \notag \\
    &= -(S_0+\hat{S_i}) M \notag \\
H_{i}^\pm &= -S_0 M \mp \Delta M.
\end{align}
In the subsequent derivation, we use no Boltzmann constant (and no $\beta$ either), just $T$.  The average value of a spin, $\bar{S_i}$, according to the Boltzmann distribution, with $P(S_{i}^-)$ standing for the probability that $S_{i}$ is negative and $P(S_{i}^+)$ that it is positive, is
\begin{align}
    \bar{S_i}   &= S^-P(S_{i}^-) + S^+P(S_{i}^+)\\
                &= \dfrac{S^- e^{-H_i^-/T} + S^+ e^{-H_i^+/T}}{e^{-H_i^-/T}+e^{-H_i^+/T}} \notag \\
                &= \dfrac{S^- e^{-(-S_0 M + \Delta M)/T} + S^+ e^{-(-S_0 M - \Delta M)/T}}{e^{(-S_0 M + \Delta M)/T}+e^{-(-S_0 M - \Delta M)/T}} \notag \\
                &= \dfrac{S^- e^{-\Delta M/T} + S^+ e^{\Delta M/T}}{e^{-\Delta M/T}+e^{\Delta M/T}} \notag \\
                &= \dfrac{(S_0 - \Delta) e^{-\Delta M/T} + (S_0 + \Delta) e^{\Delta M/T}}{e^{- \Delta M/T}+e^{\Delta M/T}} \notag \\
                &= S_0 \dfrac{e^{-\Delta M/T} + e^{\Delta M/T}}{e^{-\Delta M/T}+e^{\Delta M/T}}  +  \Delta \dfrac{-e^{-\Delta M/T} + e^{\Delta M/T}}{e^{-\Delta M/T}+e^{\Delta M/T}} \notag \\
                 &= S_0  +  \Delta \tanh{(\Delta M/T)}.  \label{eq:MF}
\end{align}

This result without non-fighters was inferred in an earlier study (author(s)). Here, we also deal with non-fighters in proportion $p > 0$. Accordingly, we define $M_{cc}$ as the average spin value of the conditional cooperators and $M_{ud}$ as the average spin value of the non-fighters. Note that $M_{ud} = S^-$. We assume that the non-fighters are homogeneously distributed across the network. Accordingly, the mean field equation becomes
\begin{align}
    S_i &= p S^- + ( 1 - p ) M_{cc}. \label{eq:pq}  
\end{align}
The Hamiltonian for a single conditional cooperator becomes
\begin{align}
H_{i}^\pm &= - ( S_0 \pm \Delta ) ( p S^- + (1 - p) M_{cc})  \\
&= - S_0 p S^- \mp \Delta p S^- -S_0 (1 - p) M_{cc} \mp \Delta (1 - p) M_{cc}. \label{eq:redundant}
\end{align}
In the derivation of Eq.~\ref{eq:MF}, all terms that did not contain $\mp \Delta$ canceled each other out. For clarity, we remove these terms from Eq.~\ref{eq:redundant}, which results in  
\begin{align}
H_{i}^\pm &= \mp \Delta (p S^- + (1 - p) M_{cc}).
\end{align}
The mean field analysis for conditional cooperator $i$ is
\begin{align}
    \bar{S_i}   &= S^-P(S_{i}^-) + S^+P(S_{i}^+)  \notag  \\
                &= \dfrac{S^- e^{- \Delta (p S^- + (1 - p) M_{cc})/T} + S^+ e^{\Delta (p S^- + (1 - p) M_{cc})/T}}{e^{-\Delta (p S^- + (1 - p) M_{cc})/T}+e^{\Delta (p S^- + (1 - p) M_{cc})/T}} \notag  \\
                &= S_0 + \Delta \dfrac{- e^{-\Delta (p S^- + (1 - p) M_{cc})/T} + e^{  \Delta (p S^- + (1 - p) M_{cc})/T}}{e^{-\Delta (p S^- + (1 - p) M_{cc})/T}+e^{\Delta (p S^- + (1 - p) M_{cc})/T}}  \notag  \\              
                &= S_0 + \Delta \tanh{(\Delta (p S^- + (1 - p) M_{cc})/T)} = M_{cc}. \label{eq:mfmcc}
\end{align}
Using Eq.~\ref{eq:pq}, 
we can express the self-consistency equation of $M_{cc}$ in $M$,
\begin{align}
M &= p S^- + (1-p)(S_0 + \Delta \tanh{(\Delta (p S^- + (1 - p) M_{cc})/T)}) 
\notag  \\
     &=  p S^- + (1-p) (S_0 + \Delta \tanh{(\Delta M/T)}) \nonumber 
\notag  \\
     &=  S_0 - p\Delta + (1-p)\Delta \tanh{(\Delta M/T)}. \label{eq:selfconsistent}
\end{align}

\subsection*{Critical proportion of non-fighters}
\begin{figure}
\begin{center}
\includegraphics[width=.7\textwidth]{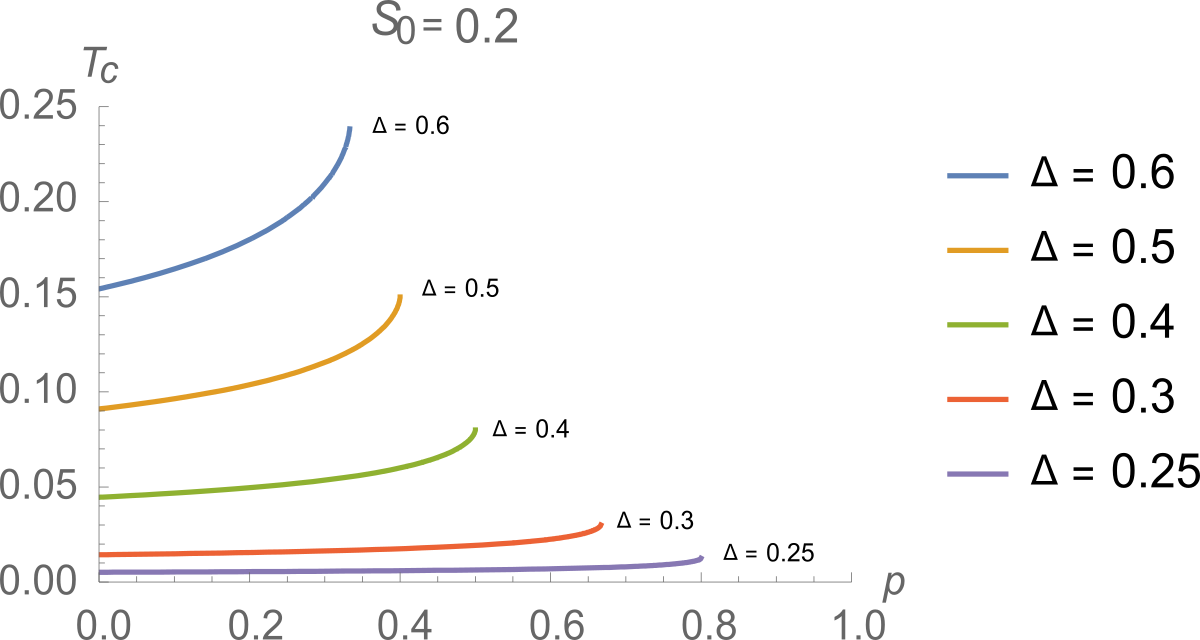} 
\end{center}
\caption{Given $S_0 = 0.2$, the critical level of agitation, $T_c$, is plotted as a function of the proportion of non-fighters, $p$, for various levels of $\Delta$. The critical level, $p_c$, is reached at the right-hand end of the lines. } 
\label{fig:S9}
\end{figure}

Depending on $T$, the self-consistency equation has one unstable and two stable ferromagnetic solutions, or one stable paramagnetic solution. 
At a critical $T$, the system transitions between these two states. When the system is paramagnetic,  $$\frac{ \partial  }{\partial M} (S_0 - p\Delta + (1-p)\Delta \tanh{(\Delta M/T)}) < 1$$ at the solution of $M$. When the system is ferromagnetic, $$\frac{ \partial  }{\partial M} (S_0 - p\Delta + (1-p)\Delta \tanh{(\Delta M/T)}) > 1$$ at the unstable solution of $M$. 
We can identify a critical $T$ when 
\begin{equation}
    \frac{ \partial  }{\partial M} (S_0 - p\Delta + (1-p)\Delta \tanh{(\Delta M/T)}) = 1, 
\end{equation}
hence,
\begin{align}
    \frac{1}{\cosh^2{(\Delta M/T)}}\Delta^2(1-p)/T &= 1
\notag \\
    \cosh{(\Delta M/T)}  &= \sqrt{\Delta^2(1-p)/T} 
\notag  \\
    \Delta M/T &= \pm  \arcosh{(\sqrt{\Delta^2(1-p)/T})} 
\notag  \\
    \frac{M}{\Delta} &= \frac{\pm \arcosh{(\sqrt{ \Delta^2(1-p)/T})}}{\Delta^2/T}.  \label{eq:Psi} 
\end{align}
We can substitute this expression (\ref{eq:Psi}) in the self-consistency equation (\ref{eq:selfconsistent})
\begin{align}
\frac{\pm \arcosh{(\sqrt{\Delta^2 (1-p)/T})}}{\Delta/T} &= S_0 - p\Delta + (1-p) \Delta \nonumber \tanh{({\pm \arcosh{( \sqrt{\Delta^2 (1-p)/T})}})} 
\notag  \\
 \frac{\pm \arcosh{(\sqrt{\Delta^2 (1-p)/T})}}{\Delta^2/T} &=  \frac{S_0}{\Delta} - p \pm (1-p) \tanh{({\arcosh{( \sqrt{\Delta^2 (1-p)/T})}})} 
\notag  \\
   &=  \frac{S_0}{\Delta} - p \pm (1-p) \sqrt{1- \frac{1}{\Delta^2(1-p)/T}}. 
\end{align}
Discarding the equation that has no real numerical solutions leaves
\begin{equation}
- \frac{\arcosh(\sqrt{\Delta^2(1-p)/T})}{\Delta^2/T} +  
(1-p) \sqrt{1- \frac{1}{\Delta^2(1-p)/T}} =  \frac{S_0}{\Delta} - p.
\end{equation}
Solutions of this equation become complex if $p >  \frac{S_0}{\Delta}$, hence $p \leq  \frac{S_0}{\Delta}$.  Graphical illustrations are in Fig.~S\ref{fig:S9}.  The choice of $C = 1$ and $D = 1/2$ implies that there is no (burst of) cooperation if $p > 1/3$.  

\subsection*{Solving the Ising model computationally}
The Ising model was solved computationally, in Fortran for speed and in R for comfort.  Surprisingly, $p_c$ of simulated, fully connected networks is very close to the mean field calculation of $p_c$, even in very small networks. Given $C = 1$ and $D = 1/2$, there is no burst if $p_c > 0.34$.  The critical threshold is nearly density independent: if in (sufficiently large) simulated networks, density is decreased by two orders of magnitude, $p_c$ increases from $\approx 0.34$ to $\approx 0.35$. The degree distribution has no effect on $p_c$. 

\begin{figure}[!ht]
\begin{center}
\includegraphics[width=.50\textwidth]{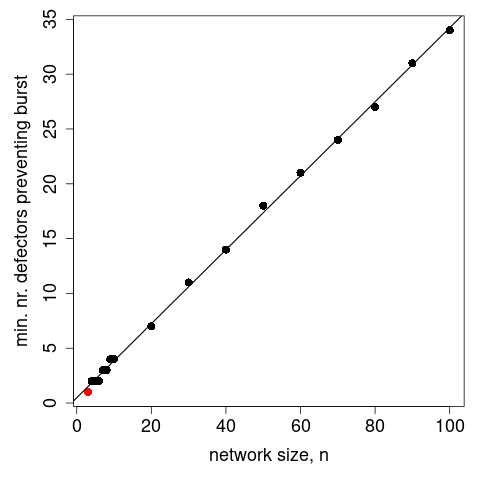} 
\end{center}
\caption{Color online. For simulated networks with sizes $3 \le n \le 100$, the minimum number of non-fighters who prevent a burst is plotted. The red dot (left, closest to the bottom) marks the triad (1 defector prevents a burst of 2), which conflicts with the empirical data (1 defector and a burst of 2).} 
\label{fig:S10}
\end{figure}

In simulations, $p_c$ is less precise in smaller networks (Table~S3) and the prediction for the triad is empirically false, whereas the mean field is correct: one non-fighter does not prevent cooperation of the other two. Also one empirical group of six with two non-fighters had a burst, which did not happen in the simulations (Fig~\ref{fig:S10}).

\begin{table}[ht!]
  \centering
     \begin{tabular}{|c|c | c | c | c|}
       \hline
       $n$ & $N_D$ & $ \approx N_C $ & burst \\
       \hline
       2 & 1 & 0 & 0 \\
       3 & 1 & 1 & 0 \\
       4 & 1 & 2 & 1  \\
       5 & 1 & 3-4 & 1 \\
       6 & 1 & 4-5 & 1 \\
       5 & 2 & 1 & 0 \\
       6 & 2 & 2 & 0 \\
       7 & 2 & 4 & 1 \\
       8 & 2 & 4-5 & 1 \\
       9 & 2 & 6 & 1 \\
       9 & 3 & 3 & 0 \\
\hline
\end{tabular}
\caption{Simulations just below (burst = 1) and above (burst = 0) the critical threshold of non-fighters ($N_D$) in small groups (size $n$). The numbers of cooperators ($N_C$) fluctuate across simulation runs and are approximate. Note: For groups with $n=7$ and $N_D=2$, simulations in Fortran yield $N_C \approx 4$ whereas in $R$, $N_C \approx 3$.}
\label{table:defect}
\end{table}

\footnotesize
\bibliographystyle{plain}
\bibliography{evo1}

\end{document}